\documentclass[
 aip,
 jcp,
 amsmath,amssymb,
 groupedaddress, 
 reprint,
]{revtex4-1}

\usepackage{graphicx}
\usepackage{color}
\usepackage{dcolumn}
\usepackage{multirow}

\begin{document}

\title[Quantum System Partitioning at the Single-Particle Level]{Quantum System Partitioning at the Single-Particle Level}

\author{Adrian H. M\"uhlbach}%
\affiliation{ETH Z\"urich, Laboratorium f\"ur Physikalische Chemie, \\ Vladimir-Prelog-Weg 2, CH-8093 Z\"urich, Switzerland
}
\author{Markus Reiher}
 \email{markus.reiher@phys.chem.ethz.ch (corresponding author)}
\affiliation{ETH Z\"urich, Laboratorium f\"ur Physikalische Chemie, \\ Vladimir-Prelog-Weg 2, CH-8093 Z\"urich, Switzerland
}

\date{17 October 2018}

\setlength{\parindent}{0cm}
\setlength{\parskip}{0.6em plus0.2em minus0.1em}

\newcommand{\farg}[1]{{\left[{\scriptstyle #1}\right]}}

\begin{abstract}
We discuss the partitioning of a quantum system by subsystem separation through unitary block-diagonalization (SSUB) applied to a Fock operator.
For a one-particle Hilbert space, this separation can be formulated in
a very general way. Therefore, it can be applied to very different partitionings ranging from those driven by features in the molecular structure 
(such as a solute surrounded by solvent molecules or an active site in an enzyme) to those that aim at an orbital
separation (such as core-valence separation). Our framework embraces recent developments
of Manby and Miller as well as older ones of Huzinaga and Cantu. Projector-based embedding is 
simplified and accelerated by SSUB. 
Moreover, it directly relates
to decoupling approaches for relativistic four-component many-electron theory. For a Fock operator
based on the Dirac one-electron Hamiltonian,  one would like to separate the so-called positronic (negative-energy) states from
the electronic bound and continuum states. The exact two-component (X2C) approach developed
for this purpose becomes a special case of the general SSUB framework and
may therefore be viewed as a system-environment decoupling approach.
Moreover, for SSUB there exists no restriction with respect to the number of subsystems that are generated --- in the limit,
decoupling of all single-particle states is recovered, which represents exact diagonalization of the problem.
The fact that a Fock operator depends on its eigenvectors poses challenges to all
system-environment decoupling approaches and is discussed in terms of the SSUB framework.
Apart from improved conceptual understanding, these relations bring about technical advances
as developments in different fields can immediately cross-fertilize
one another. As an important example we discuss the atomic decomposition of the unitary block-diagonalization matrix in X2C-type
approaches that can inspire approaches for the efficient partitioning of large total systems based on SSUB.
\end{abstract}

\maketitle

\section{Introduction}
The quantum mechanical study of isolated molecular systems has been an important endeavor.
Examples range from scrutinizing our understanding of fundamental physical theory 
(as highlighted, for instance, by the high resolution
results available for the dihydrogen binding energy
\cite{Cheng18,PhysRevA.97.060501,puch18})
to analyzing vast amounts of experimental (gas-phase) data in great detail
(examples can be found in astrochemistry \cite{Barone15,Puzzarini18}
as well as in atmospheric and combustion chemistry \cite{Glowacki12}).
However, the majority of experiments in chemistry considers molecules in some specific environment
(in solution, on surfaces, in solid bulk, in enzymes and so forth),
which poses huge challenges for their theoretical description.

Naturally, a plethora of approximations has been developed to cope with situations in which a local phenomenon, i.e.,
one that can be described by studying only a subsystem, is embedded into some environment that more or less strongly interacts
with the subsystem.
Some of these embedding approaches were driven by chemical and physical insights resting
on rather ad hoc theoretical bases of which quantum-mechanics molecular-mechanics (QMMM) coupling \cite{Warshel76,Singh86,Field90,Lin06,Senn07b,Senn07a,Senn09}
is the most prominent example including its sophisticated variants such as polarizable embedding theories \cite{Olsen10,Olsen11,Sneskov11}. 
Various fragmentation and embedding approaches
were conceived to enhance computational efficiency by reducing the number of one-particle basis
functions or by fragmenting the system, which also make calculations amenable
to massive parallelization; examples can be found in Refs.\ \onlinecite{Svensson96, Dapprich99, Nakai02, Slipchenko06, Gordon07, Akama07, Kobayashi09, Gordon12}.

From the more formal point of view of quantum theory, nesting a subsystem into an environment of one or
more subsystems requires the adoption of open-system quantum mechanics, \cite{Breuer02,Amann11}
which in principle can cope with any such situation. For an open quantum system, many-particle
basis states defined on a subsystem may not necessarily conserve particle number as they can be combined with states
from the environment to produce a total state of, in most practical cases,
fixed particle number. The total state may then be expanded in terms of a (tensor) product basis where the double sum runs over
indices that refer to subsystem (sub)states and to environment (sub)states.
Such a partitioning of a system can be directly exploited to optimize basis states on a subsystem
in numerical procedures. The density matrix renormalization group algorithm \cite{White92,White93}
is an example, where in each iteration step a total many-particle state may be viewed as being decomposed into a  
product basis of substates defined on a system and an environment of orbitals. 

A very special decomposition is the Schmidt decomposition\cite{Schmidt07,Schollwoeck11}, 
which restricts the double sum over product states to a single sum by connecting
each state on a system to exactly one (specially prepared, e.g., contracted) many-particle state of the environment. It 
is this decomposition that has prompted Knizia and Chan
to define an efficient embedding model called 
density matrix embedding theory (DMET)\cite{Knizia12,Knizia13}. DMET exploits the fact that a potentially small number of relevant system
states couples, by virtue of the Schmidt decomposition,
to only the same number of states in the environment, no matter
how large the latter is. Obviously, the optimization of such environment states might be considered as complicated
as solving the full quantum problem for the total system (i.e., for subsystem and environment).
To arrive at a practical DMET approach, Knizia and Chan proposed a mean-field approximation for the environment states.\cite{Knizia12,Knizia13}
The mean-field approximation to the general DMET has been studied in detail by them,\cite{Knizia12,Knizia13}
by Scuseria and co-workers\cite{Bulik14,Bulik14a}, and by van Voorhis and co-workers\cite{Welborn16,Ricke17}.

Mean-field environments had been considered for system-environment partitioning before the introduction of DMET. The motivation for this has always
been the observation that a part of a total system may be subject to strong quantum correlations whereas for
the rest a mean-field approach can be chosen, which is usually taken to be Kohn--Sham density functional
theory (KS-DFT)\cite{Hohenberg64,Kohn65}. Within DFT, it is possible to define density-based formulations of a system-environment embedding.
\cite{Gordon72, Kim74, Senatore86, Cortona91, Wesolowski93, Cortona94, Neugebauer05, Iannuzzi06, Jacob08, Fux10, Elliott10, Goodpaster10, Jacob14, Fornace15, Wesolowski15, Ding17}
The strongly correlated part of a molecule, i.e., the system, may also be described by an accurate
wave-function-theory approach
\cite{Huzinaga71, Govind99, Kluener02, Huang06, Gomes08, Huang11, Manby12, Hoefener12, Goodpaster14, Daday14, Dresselhaus15, Hegely16} if deemed necessary to allow for better error control.

Mean-field approximations lend themselves to studying quantum system partitioning at the single-particle level, i.e., at the level
of the one-particle equations of motion that describe the dynamics of an electron in a mean-field potential. Obviously, Hartree--Fock
and Kohn--Sham equations are the most popular targets for such a decomposition. 
In this work, we will present a general unitary-transformation-based 
partitioning approach for single-particle equations.
We would like to emphasize, however, that these single-particle equations do not need to be of the mean-field type. Our unitary decoupling
approach will apply to any single-particle equation, which could, for instance, be of a multi-configuration self-consistent-field type,
in which configuration-interaction state parameters enter the electron-electron interaction at the one-particle level.

We briefly mention that other embedding theories exploit different formulations of the quantum mechanical equations of motion.
Examples are the self-energy embedding theory of Zgid that starts from a Green's function formalism\cite{Kananenka15,Lan15,Lan17},
the dynamical mean-field theory of Georges and Kotliar for the description of impurities\cite{Georges92,Georges96,Georges04,Kotliar06},
and work that allows one to nest different quantum formalisms into one another \cite{Fromager15,Senjean17,Senjean18}.
Also active-orbital space methods \cite{Roos80,Roos80a,Ruedenberg82,Olsen88,Fleig03,Ivanic03}
can be viewed as embedding approaches nesting a set of strongly statically correlated orbitals considered for
exact diagonalization into the complementary space of all other less correlated orbitals,
as recently exploited by Shiozaki and co-workers in what they call the active space decomposition method\cite{Parker13,Parker14,Parker14a}.

Whereas all general open-quantum-system methods operate, as they should, on the many-particle state level,
such separations of a system into subsystems can be leveraged by a properly prepared
one-particle basis from which the many-particle states are then constructed (either in the
usual way by tensor products or in a mean-field sense for KS-DFT).
This was recently demonstrated in the work of Manby, Miller, and co-workers
on different variants of embedding\cite{Manby12, Goodpaster14, Fornace15} 
and we come back to their embedding approaches later in this work.

In this work, we consider the separation of a quantum system by a suitable linear combination of one-electron basis states
that allows us to provide a separation according to any desired target, which may be defined in terms of an underlying nuclear
framework or by exploiting a separation of one-particle states based on some energy criterion (producing, for instance, core-valence separation
within an atomic or molecular structure). Our approach is designed to be efficient and generally
applicable. It even relates to exact two-component
relativistic theories. 
However, the fact that a Fock operator usually depends on the solution of a one-particle equation 
(as, in general, the 4-current or the electron density (matrix)
is required to represent the interaction) will pose difficulties for all such embedding approaches, but at the same time, facilitates the
proposition of suitable approximations. Within our general framework, we will show that approximations developed 
for exact two-component approaches may cross-fertilize developments in
embedding theories by Miller and Manby.

\section{One-Electron Hilbert space}\label{sec:MF}
We discuss the partitioning of a system at the level of a one-electron equation and assume that one can construct a Fock matrix for the 
total system and eventually diagonalize it. Clearly, if the system is very large, this will become a problem. For such cases, it will be 
necessary to introduce focused methods to construct and diagonalize Fock matrices, which are, 
for instance, known in plane-wave calculations. 
QMMM may be viewed as a radical solution that treats part of a system
classical so that it does not at all contribute any basis states for the representation of the Fock operator, whereas semiempirical methods\cite{Thiel88, Dewar92, Clark93, Thiel96, Thiel98, Clark00, Bredow05, Stewart07, Lewars10, Clark11, Thiel14, Bredow17, Husch18}
allow one to approximate fairly large Fock matrices. 

For the sake of brevity, we focus on mean-field equations
and consider the restricted formalism only. 
An extension to an unrestricted formalism is straightforward. 
We first give a concise overview on the general formalism to introduce a unified notation and keep our account self-contained.

The Fock matrix $\mathbf{F}$ is the representation of a Fock operator $\hat{F}$ within a basis $B$ consisting of $N_B$ basis functions,
\begin{align}
B = \{\phi_i: i \in [1; N_B]\}.
\end{align}
The Fock matrix $\mathbf{F}$ depends on the density matrix $\mathbf{P}$, $\mathbf{F}\farg{\mathbf{P}}$.
It is the sum of the one-electron matrix $\mathbf{H}$, which does not depend on the density matrix $\mathbf{P}$, and the two-electron matrix $\mathbf{V} = \mathbf{V}\farg{\mathbf{P}}$,
\begin{align}
\mathbf{F}\farg{\mathbf{P}} = \mathbf{H} + \mathbf{V}\farg{\mathbf{P}},\label{eq:F_MF}
\end{align}
both of which are hermitian.
The (closed-shell) density matrix $\mathbf{P}$,
\begin{align}
\mathbf{P} = 2 \, \mathbf{C}_\text{occ} \left(\mathbf{C}_\text{occ}\right)^\mathsf{T}, \label{eq:P_def}
\end{align}
is calculated from the molecular orbitals $\psi_i^\text{occ}$, which occupy the Hartree--Fock (HF) Slater determinant,
in the chosen basis,
\begin{align}
\psi_i^\text{occ} = \sum_{k=1}^{N_B} c^{(k)}_{i_\text{occ}} \phi_k
\end{align}
where the $c^{(k)}_{i_\text{occ}}$ are elements of a vector $c_{i_\text{occ}}$ representing
the $i$-th occupied orbital in this basis. All vectors $c_{i_\text{occ}}$ enter $\mathbf{C}_\text{occ}$ as column vectors.
Their determination requires diagonalization of the Fock matrix $\mathbf{F}$.

The one-electron matrix $\mathbf{H}$ consists of contributions from the kinetic energy of an electron and the potential energy arising from the attractive Coulomb interaction between an electron and the $N_\text{nuc}$ nuclei of the system. 
We write its matrix elements $H_{ij}$ in Hartree atomic units (used throughout) as
\begin{align}
H_{ij} = -\frac{1}{2}\langle\phi_i|\Delta|\phi_j\rangle - \sum_{I}^{N_\text{nuc}} Z_I \langle\phi_i|\frac{1}{r_I}|\phi_j\rangle,\label{eq:H_MF}
\end{align}
with the Laplace operator in three dimensions $\Delta$,
the nuclear charge number of the $I$-th nucleus $Z_I$,
and the Euclidean distance $r_I$ 
between the integration coordinate and the position of nucleus $I$.

In a general framework that considers Hartree--Fock and Kohn--Sham density functional theory on the same 
algorithmic footing, the two-electron matrix $\mathbf{V}\farg{\mathbf{P}}$ may be thought of as consisting of the two-electron Coulomb matrix $\mathbf{J}\farg{\mathbf{P}}$, the two-electron exchange matrix $\mathbf{K}\farg{\mathbf{P}}$ and the Kohn--Sham exchange-correlation matrix $\mathbf{V}_\text{xc}\farg{\mathbf{P}}$,
\begin{align}
\mathbf{V}\farg{\mathbf{P}} = \mathbf{J}\farg{\mathbf{P}} + \alpha \mathbf{K}\farg{\mathbf{P}} + \beta \mathbf{V}_\text{xc}\farg{\mathbf{P}}, \label{eq:V_MF}
\end{align}
where $\alpha$ mixes in exact (Hartree--Fock) exchange that needs to be corrected for in $\mathbf{V}_\text{xc}$ (not shown).
In Hartree--Fock theory, $\alpha$=1 and $\beta$=0 in Eq.~\eqref{eq:V_MF}.
In Kohn--Sham density functional theory, $\beta$=1 and $\alpha$ controls the amount of exact exchange admixture.

Each component of $\mathbf{V}$ depends on the one-electron density matrix $\mathbf{P}$.
The elements of the Coulomb matrix $\mathbf{J}\farg{\mathbf{P}}$ and the exchange matrix $\mathbf{K}\farg{\mathbf{P}}$ are calculated from 
the two-electron repulsion integrals evaluated in the chosen basis as
\begin{align}
J_{ij}\farg{\mathbf{P}} = \sum^{N_B}_{kl} P_{kl} \langle\phi_i(1)\phi_k(2)|\frac{1}{r_{12}}|\phi_j(1)\phi_l(2)\rangle\label{eq:J_MF}
\end{align}
and
\begin{align}
K_{ij}\farg{\mathbf{P}} = -\frac{1}{2}\sum^{N_B}_{kl} P_{kl} \langle\phi_i(1)\phi_k(2)|\frac{1}{r_{12}}|\phi_l(1)\phi_j(2)\rangle,\label{eq:K_MF}
\end{align}
respectively. Here, $r_{12}$ denotes the Euclidean distance between the two integration coordinates.
In Kohn--Sham density functional theory \cite{Hohenberg64, Kohn65}, the exchange-correlation matrix $\mathbf{V}_\text{xc}\farg{\mathbf{P}}$ is calculated from the exchange-correlation potential $v_\text{xc}{\left[\rho\farg{\mathbf{P}}\right]}$,
\begin{align}
V_{\text{xc},ij}\farg{\mathbf{P}} = \langle\phi_i|v_\text{xc}{\left[\rho\farg{\mathbf{P}}\right]}|\phi_j\rangle, \label{eq:VXC_MF}
\end{align}
with the electron density $\rho\farg{\mathbf{P}}$, 
\begin{align}
\rho\farg{\mathbf{P}} = \sum_{ij}^{N_B} P_{ij} \phi_i \phi_j, \label{eq:RHO_MF}
\end{align}
where we assume real orbitals.

The electronic energy $E_\text{el}\farg{\mathbf{P}}$ is the sum of the Coulomb energy of all nuclei $E_\text{nuc}$, the one-electron energy $E_H\farg{\mathbf{P}}$, the Coulomb and exchange energies 
$E_{JK}\farg{\alpha,\mathbf{P}}$, and the exchange-correlation functional $E_\text{xc}\farg{\mathbf{P}}$,
\begin{align}
E_\text{el}\farg{\mathbf{P}} = E_\text{nuc} + E_H\farg{\mathbf{P}} + E_{JK}\farg{\alpha,\mathbf{P}} + \beta E_\text{xc}\farg{\mathbf{P}}.
\end{align}
The Coulomb energy of all nuclei is calculated as
\begin{align}
E_\text{nuc} = \sum^{N_\text{nuc}}_{i<j} \frac{Z_i Z_j}{r_{ij}}.
\end{align}
The one-electron energy $E_H\farg{\mathbf{P}}$ collects the density matrix weighted contributions from the corresponding one-electron matrix $\mathbf{H}$,
\begin{align}
E_H\farg{\mathbf{P}} = \sum^{N_B}_{ij} H_{ij} P_{ji} = \text{Tr}{\left(\mathbf{H}\mathbf{P}\right)}.
\end{align}
The energy contribution $E_{JK}\farg{\alpha,\mathbf{P}}$ is evaluated from the Coulomb and 
exchange matrices $\mathbf{J}$ and $\mathbf{K}$, respectively,
\begin{align}
E_{JK}\farg{\mathbf{P}} = \frac{1}{2} \sum^{N_B}_{ij} \left(J_{ij} + \alpha K_{ij}\right) P_{ji} = \frac{1}{2} \text{Tr}{\left(\left(\mathbf{J} + \alpha \mathbf{K}\right)\mathbf{P}\right)}.
\end{align}

\section{The self-consistent-field procedure}
To obtain the ground-state energy, the electronic energy is minimized in a self-consistent manner
through the self-consistent field (SCF) procedure.
The following description will only consider a basic formulation that is required to
later discuss all options of self-consistent and approximate embedding schemes.

Quantities that do not depend on the density matrix $\mathbf{P}$ are evaluated before the iterative part of the SCF procedure starts. 
Apart from the one-electron matrix $\mathbf{H}$ whose elements were defined in Eq.~\eqref{eq:H_MF}, this is 
the overlap matrix $\mathbf{S}$, with elements
\begin{align}
S_{ij} = \langle\phi_i|\phi_j\rangle
\end{align}
Depending on hardware constraints, the two-electron repulsion integrals in the atomic-orbital basis,
which are required for the evaluation 
of the Coulomb and exchange matrices in Eqs.~\eqref{eq:J_MF} and \eqref{eq:K_MF}, may be precalculated
as well.
As a starting point for the minimization, an initial density matrix $\mathbf{P}^{(0)}$ is to be determined, 
for which various options exist (e.g., superposition of atomic densities, extended H\"uckel theory guess, 
basis set projection, or the diagonalization of the one-electron matrix $\mathbf{H}$). 
Then, in the $n$-th iteration step, 
the Fock matrix $\mathbf{F}^{(n)} = \mathbf{F}\farg{\mathbf{P}^{(n-1)}}$ is calculated from the 
density matrix of the previous step, $\mathbf{P}^{(n-1)}$.

The generalized eigenvalue problem of the Roothaan--Hall equation\cite{Roothaan51,Hall51} is solved 
to obtain the $n$-th approximation to the eigenvector matrix $\mathbf{C}^{(n)}$ and 
to the diagonal matrix of eigenvalues $\boldsymbol{\epsilon}^{(n)}$,
\begin{align}
\mathbf{F}^{(n)}\mathbf{C}^{(n)} = \mathbf{S}\mathbf{C}^{(n)}\boldsymbol{\epsilon}^{(n)}. \label{eq:RHE}
\end{align}
This can be achieved by converting it to the ordinary eigenvalue problem
\begin{align}
\mathbf{S}^{-1}\mathbf{F}^{(n)}\mathbf{C}^{(n)} = \mathbf{C}^{(n)}\boldsymbol{\epsilon}^{(n)}.
\end{align}

The molecular orbitals are defined by the eigenvectors in the atomic-orbital basis $B$ chosen,
\begin{align}
\xi_i^{(n)} = \sum_{j=1}^{N_B} c_{i,j}^{(n)} \phi_j = \sum_{j=1}^{N_B} \mathbf{C}_{ij}^{(n)} \phi_j.
\end{align}
The orbital energy of the molecular orbital $\xi_i^{(n)}$ corresponds to the diagonal entry $\epsilon_{ii}^{(n)}$ in the eigenvalue matrix $\boldsymbol{\epsilon}^{(n)}$.
The $N_\text{occ}$ molecular orbitals with the lowest orbital energies enter the Hartree--Fock
determinant (in the restricted formalism, $N_\text{occ}$ is equal to half the number of electrons $N_\text{el}$).
From the matrix $\mathbf{C}_\text{occ}^{(n)}$ of occupied eigenvectors,
\begin{align}
\mathbf{C}^{(n)}_\text{occ} = 
\begin{pmatrix}
c_{i_1}^{(n)} & c_{i_2}^{(n)} & \cdots & c_{i_{N_\text{occ}}}^{(n)}
\end{pmatrix},
\end{align}
a new density matrix $\mathbf{P}^{(n)}$, 
\begin{align}
\mathbf{P}^{(n)} = 2 \, \mathbf{C}^{(n)}_\text{occ} \left(\mathbf{C}^{(n)}_\text{occ}\right)^\mathsf{T}, \label{eq:P_construction}
\end{align}
is calculated.

As convergence criteria can serve the Frobenius norm of the difference between two consecutive density matrices, 
\begin{align}
\delta_\mathbf{P}^{(n)} = \left\|\mathbf{P}^{(n-1)} - \mathbf{P}^{(n)}\right\|, \label{eq:D_P}
\end{align} 
and between total electronic energies calculated from them,
\begin{align}
\delta_{E_\text{el}}^{(n)} = \left| E_\text{el}\farg{\mathbf{P}^{(n-1)}} - E_\text{el}\farg{\mathbf{P}^{(n)}} \right|, \label{eq:D_E}
\end{align}
to be below a predefined threshold. 
If the procedure has not yet converged, a new iteration step will begin where the new density matrix $\mathbf{P}^{(n)}$ is injected into the calculation of the next Fock matrix $\mathbf{F}^{(n+1)} = \mathbf{F}\farg{\mathbf{P}^{(n)}}$.

\begin{figure*}[htbp!] 
\includegraphics[width=\textwidth]{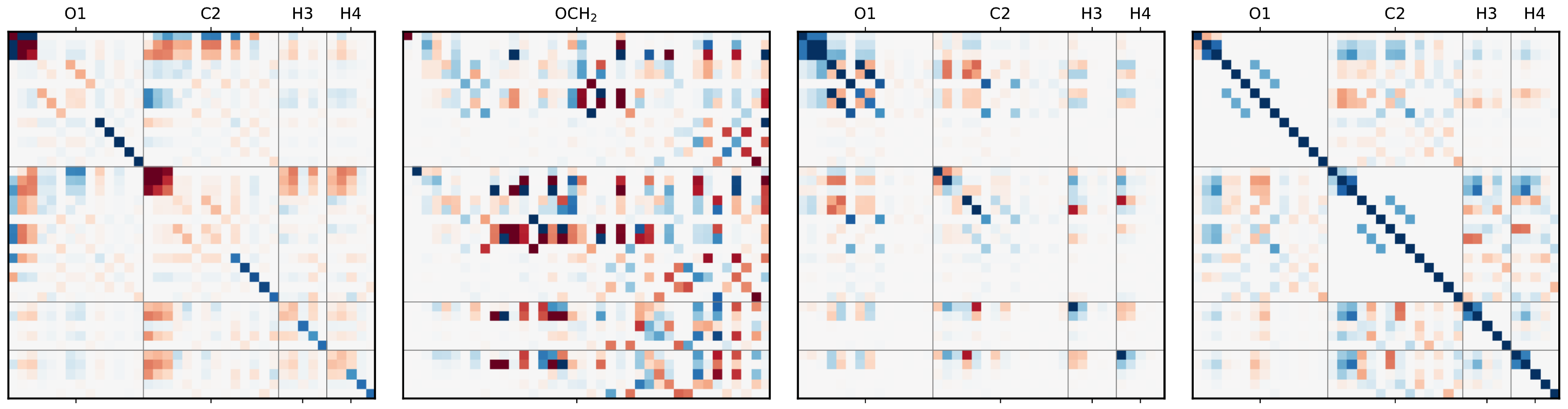}
\caption{Fock matrix $\mathbf{F}$, eigenvector matrix $\mathbf{C}$, density matrix $\mathbf{P}$, and overlap matrix $\mathbf{S}$ (from left to right) of a converged Hartree--Fock calculation for formaldehyde. 
The color code is such that red indicates negative values, whereas blue stands for positive ones.
Then, the matrix-element values covered by the color range from red to blue for the different matrices are as follows:
$\mathbf{F}$ $-$2 to +2,
$\mathbf{C}$ $-$1 to +1,
$\mathbf{P}$ $-$0.4 to +0.4, and
$\mathbf{S}$ from $-$1 to +1.
Values beyond these ranges are marked in deep red or deep blue.
\label{fig:FCPS}}
\end{figure*}

We implemented all procedures discussed in this paper into a local version of PySCF 1.5b\cite{Sun18}. 
The Def2-SVP\cite{Weigend05} basis set was chosen for all calculations carried out in this work, which are all carried out in the Hartree--Fock approximation
(all molecular structures are provided in the supporting information). At the example of
formaldehyde, Fig.~\ref{fig:FCPS} shows
the structure of the Fock matrix $\mathbf{F}$, eigenvector matrix $\mathbf{C}$, density matrix $\mathbf{P}$, and overlap matrix $\mathbf{S}$,
to which we later compare transformed matrices emerging in the embedding approaches.

\section{Fock matrix block-diagonalization}\label{sec:FBD}
In this section, we introduce a general decoupling approach that eliminates interactions between subsystems of a molecular system at the matrix level. 
This then allows for separate treatments of subsystems.
Decoupling is achieved through two subsequent matrix transformations: first an orthogonalization of the basis, then a block-diagonalization of the orthogonalized Fock matrix.
In comparison with orbital localization schemes, which prepare the one-electron basis and then evaluate the operator matrices in this
tailored basis, we manipulate the operator matrices to achieve exact block-diagonalization and 
then obtain basis states that are localized on either subsystem, but not 
necessarily localized within this subsystem.

\subsection{Transforming the basis}\label{sec:change_of_basis}
We first elaborate on the implications of a modified basis for the SCF procedure.
As we will consider cases where the transformation would be required
in each iteration step, we include the superscript '($n$)' for the SCF iteration steps.

A transformation matrix $\mathbf{W}$ transforms the initial basis set $B = \{\phi_i\}$ into the new basis set $\tilde{B} = \{\tilde{\phi}_i\}$,
\begin{align}
\tilde{\phi}_i = \sum_{j=1}^{N_B} W_{ij} \phi_j. \label{eq:basis_transformation}
\end{align}
A matrix representation $\mathbf{A}$ of an operator $\hat{A}$ in the new basis $\tilde{B}$ can then be expressed through the congruent transformation,
\begin{align}
\tilde{\mathbf{A}} = \mathbf{W} \mathbf{A} \mathbf{W}^\mathsf{T}. \label{eq:transformation}
\end{align}
It might be not convenient to evaluate the expressions for the transformed Fock matrix in terms of the transformed density matrix $\tilde{\mathbf{P}}$ in the basis $\tilde{B}$
as it is necessary to introduce additional transformation steps into the SCF iterations:

With a Fock matrix $\tilde{\mathbf{F}}^{(n)}$ in basis $\tilde{B}$ that 
is evaluated from the transformation of the original Fock matrix $\mathbf{F}^{(n)}$,
\begin{align}
\tilde{\mathbf{F}}^{(n)} = \mathbf{W} \mathbf{F}^{(n)} \mathbf{W}^\mathsf{T},
\end{align}
the Roothaan--Hall equation becomes
\begin{align}
\begin{split}
\mathbf{W} \mathbf{F}^{(n)} \mathbf{W}^\mathsf{T} \mathbf{W}^{-\mathsf{T}} &\mathbf{C}^{(n)} = \\
&\mathbf{W} \mathbf{S} \mathbf{W}^\mathsf{T} \mathbf{W}^{-\mathsf{T}} \mathbf{C}^{(n)} \boldsymbol{\epsilon}^{(n)},\label{eq:RHE_transform}
\end{split}
\end{align}
or expressed in terms of the transformed matrix quantities,
\begin{align}
\tilde{\mathbf{F}}^{(n)} \tilde{\mathbf{C}}^{(n)} = \tilde{\mathbf{S}} \tilde{\mathbf{C}}^{(n)} \boldsymbol{\epsilon}^{(n)},
\end{align}
with $\tilde{\mathbf{C}}^{(n)} = \mathbf{W}^{-\mathsf{T}} \mathbf{C}^{(n)}$.
Solving this new generalized eigenvalue equation yields the transformed molecular orbital coefficient matrix $\tilde{\mathbf{C}}^{(n)}$ and the eigenvalue matrix $\boldsymbol{\epsilon}^{(n)}$.

The occupied eigenvector matrix $\tilde{\mathbf{C}}^{(n)}_\text{occ}$ is constructed as before and the
transformed density matrix $\tilde{\mathbf{P}}^{(n)}$ is calculated from the transformed eigenvector matrix $\tilde{\mathbf{C}}^{(n)}$,
\begin{align}
\tilde{\mathbf{P}}^{(n)} = 2 \hspace{0.1cm} \tilde{\mathbf{C}}^{(n)}_\text{occ} \left(\tilde{\mathbf{C}}^{(n)}_\text{occ}\right)^\mathsf{T} \label{eq:P_T_construction}
\end{align}
The density matrix $\mathbf{P}^{(n)}$ in the original basis $B$ can be recovered by a back-transformation of $\tilde{\mathbf{P}}^{(n)}$,
\begin{align}
\mathbf{P}^{(n)} = \mathbf{W}^\mathsf{T} \tilde{\mathbf{P}}^{(n)} \mathbf{W}. \label{eq:p_backtransform}
\end{align}
This back-transformation is necessary because it would be inefficient to evaluate a new 
transformed Fock matrix $\tilde{\mathbf{F}}^{(n+1)}$ from the transformed density matrix $\tilde{\mathbf{P}}^{(n)}$ directly because of the 4-index transformation required for Eqs.~\eqref{eq:J_MF} and \eqref{eq:K_MF}. 
Hence, the calculation of the new transformed Fock matrix $\tilde{\mathbf{F}}^{(n+1)}$ may be more efficiently achieved by a sequence of backward and forward transformations,
\begin{align}
\tilde{\mathbf{F}}^{(n+1)} = \mathbf{W} \mathbf{F}\farg{\mathbf{W}^\mathsf{T} \tilde{\mathbf{P}}^{(n)} \mathbf{W}} \mathbf{W}^\mathsf{T}.
\end{align}

\subsection{Partitioning of the system} \label{sec:part}
In most of the following, we consider a subdivision of a molecular system into two parts, denoted \textit{subsystem} ($\mathcal{S}$) and \textit{environment} ($\mathcal{E}$).
However, an extension to an arbitrary number of subsystems (including hierarchical subsystem nesting) is also discussed.

In this work, subsystem and environment are chosen according to a partitioning of the atom-centered basis set $B$ into the subsets $B_\mathcal{S}$ and $B_\mathcal{E}$, such that
\begin{align}
B_\mathcal{S} \cup B_\mathcal{E} &= B,\\
B_\mathcal{S} \cap B_\mathcal{E} &= \varnothing.
\end{align}
The subsets $B_\mathcal{S}$ and $B_\mathcal{E}$ consist of $n_\mathcal{S}$ and $n_\mathcal{E}$ basis functions, respectively.
It is convenient to order the basis functions $\phi_i$, assigning the index $i$ such that they consist of two contiguous groups, pertaining to the subsystem and the environment.
This leads to the following definition for the subsets $B_\mathcal{S}$ and $B_\mathcal{E}$,
\begin{align}
B_\mathcal{S} &= \left\lbrace\phi_i: i \in [1; N_\mathcal{S}]\right\rbrace,\\
B_\mathcal{E} &= \left\lbrace\phi_i: i \in [N_\mathcal{S}+1; N_B]\right\rbrace.
\label{eq:basis_subsets}
\end{align}
This ordering ensures that every matrix representation $\mathbf{A}$ of an operator $\hat{A}$ can be split into distinct subblocks,
\begin{align}
\mathbf{A} = \begin{pmatrix}
\mathbf{A}_{11} & \mathbf{A}_{12}\\
\mathbf{A}_{21} & \mathbf{A}_{22}
\end{pmatrix}.
\end{align}
The subblocks $\mathbf{A}_{11}$, $\mathbf{A}_{12}$, $\mathbf{A}_{21}$, and $\mathbf{A}_{22}$ are of size $N_\mathcal{S}{\times}N_\mathcal{S}$, $N_\mathcal{S}{\times}N_\mathcal{E}$, $N_\mathcal{E}{\times}N_\mathcal{S}$, and $N_\mathcal{E}{\times}N_\mathcal{E}$, respectively. Whenever we encounter block-diagonal matrices where it is possible to assign each block to either the subsystem or the environment, we will denote the diagonal blocks as $\mathbf{A}_\mathcal{S}$ and $\mathbf{A}_\mathcal{E}$,
\begin{align}
\mathbf{A} = \begin{pmatrix}
\mathbf{A}_\mathcal{S} & \mathbf{0}\\
\mathbf{0} & \mathbf{A}_\mathcal{E}
\end{pmatrix}.
\end{align}
It should be noted that whilst we pursue an atom-wise partitioning in most of this work, it is 
possible to have any kind of partition of the basis set $B$ -- even basis functions centered on the same atom can be partitioned into both $B_\mathcal{S}$ and $B_\mathcal{E}$ if deemed useful (consider, for example, core-valence separations).
If the ordering of the basis functions differs from the specification above, it will be trivial to construct permutation matrices which will produce a block structure according to this partitioning scheme.

\subsection{Orthogonalization of the overlap matrix}\label{sec:FBD_orth}
\begin{figure*}[htbp!] 
\includegraphics[width=\textwidth]{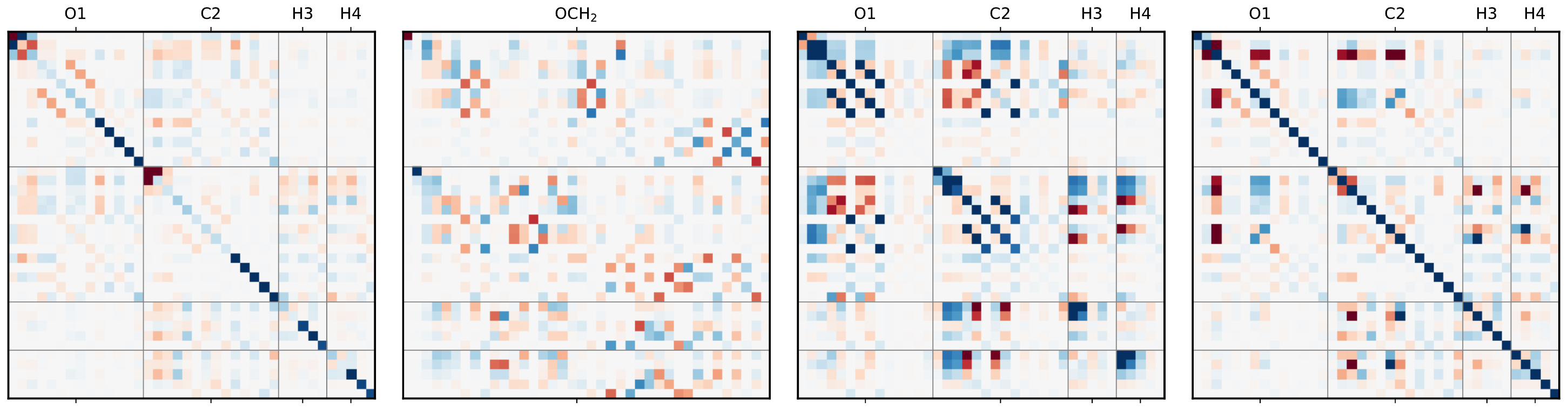}
\caption{Orthogonalized Fock matrix $\breve{\mathbf{F}}$, eigenvector matrix $\breve{\mathbf{C}}$, density matrix $\breve{\mathbf{P}}$, and orthogonalization matrix $\mathbf{X}$ (from left to right) of a Hartree--Fock calculation for formaldehyde. Color coding as in Fig.~\ref{fig:FCPS} and for
$\mathbf{X}$ we have a range of $-$1 to 1.
\label{fig:FCPS_orth}}
\end{figure*}

We now transform the generalized eigenvalue problem in Eq.~\eqref{eq:RHE} into an ordinary eigenvalue equation. This can be done with L\"owdin's symmetric orthogonalization\cite{Loewdin70}, a congruent transformation,
\begin{align}
\mathbf{X} \mathbf{S} \mathbf{X}^\mathsf{T} = \mathbf{I},
\end{align}
with
\begin{align}
\mathbf{X} = \mathbf{S}^{-\frac{1}{2}}.
\end{align}
As it is well known, we arrive at a transformed eigenvalue equation,
\begin{align}
\mathbf{X} \mathbf{F} \mathbf{X}^\mathsf{T} \mathbf{X}^{-\mathsf{T}} \mathbf{C} = \mathbf{X} \mathbf{S} \mathbf{X}^\mathsf{T} \mathbf{X}^{-\mathsf{T}} \mathbf{C} \boldsymbol{\epsilon},
\end{align}
which can be rewritten as,
\begin{align}
\breve{\mathbf{F}} \breve{\mathbf{C}} &= \breve{\mathbf{C}} \boldsymbol{\epsilon}, \label{eq:f_eig_orth}
\end{align}
with the definition of the transformed Fock matrix $\breve{\mathbf{F}} = \mathbf{X} \mathbf{F} \mathbf{X}^\mathsf{T}$ and its eigenvector matrix $\breve{\mathbf{C}} = \mathbf{S}^{\frac{1}{2}} \mathbf{C}$.
The transformed Fock matrix $\breve{\mathbf{F}}$ is hermitian.
The diagonal matrix $\boldsymbol{\epsilon}$ containing the eigenvalues is invariant under this transformation.
The structure of the matrices in Fig.~\ref{fig:FCPS} after orthogonalization is
depicted in Fig.~\ref{fig:FCPS_orth}.

\subsection{Block-diagonalization of the Fock matrix}\label{sec:FBD_bd}
At the heart of our decoupling method is a transformation matrix $\mathbf{Q}$ separating subsystem and environment at the matrix level. 
To this end, we seek to transform the Fock matrix $\breve{\mathbf{F}}$ to a basis $\tilde{B}$ in which 
it assumes a block-diagonal form,
\begin{align}
\tilde{\mathbf{F}} = \mathbf{Q} \breve{\mathbf{F}} \mathbf{Q}^\mathsf{T} = 
\begin{pmatrix}
\tilde{\mathbf{F}}_{\mathcal{S}} & \mathbf{0}\\
\mathbf{0} & \tilde{\mathbf{F}}_{\mathcal{E}}\\
\end{pmatrix}. \label{eq:BD}
\end{align}
Here, the block-diagonalization matrix $\mathbf{Q}$ needs to be unitary. 
Otherwise, the eigenvalues will not be invariant and the resulting block-diagonal Fock matrix will not be hermitian.

Such a block-diagonalization is not unique.
Given that a matrix $\mathbf{Q}$ block-diagonalizes the matrix $\breve{\mathbf{F}}$, then any matrix $\mathbf{D} \mathbf{Q}$, with a unitary block-diagonal matrix $\mathbf{D}$, also transforms $\breve{\mathbf{F}}$ into a block-diagonal form,
\begin{align}
\mathbf{D} \mathbf{Q} \breve{\mathbf{F}} (\mathbf{D} \mathbf{Q})^\mathsf{T}
&= \mathbf{D} \mathbf{Q} \breve{\mathbf{F}} \mathbf{Q}^\mathsf{T} \mathbf{D}^\mathsf{T} \\
&\hspace{-1cm}=
\begin{pmatrix}
\mathbf{D}_{11} & \mathbf{0}\\
\mathbf{0} & \mathbf{D}_{22}\\
\end{pmatrix}
\begin{pmatrix}
\tilde{\mathbf{F}}_{11} & \mathbf{0}\\
\mathbf{0} & \tilde{\mathbf{F}}_{22}\\
\end{pmatrix}
\begin{pmatrix}
\mathbf{D}^\mathsf{T}_{11} & \mathbf{0}\\
\mathbf{0} & \mathbf{D}^\mathsf{T}_{22}\\
\end{pmatrix} \\
&\hspace{-1cm}=
\begin{pmatrix}
\mathbf{D}_{11} \tilde{\mathbf{F}}_{11} \mathbf{D}^\mathsf{T}_{11} & \mathbf{0}\\
\mathbf{0} & \mathbf{D}_{22} \tilde{\mathbf{F}}_{22} \mathbf{D}^\mathsf{T}_{22}\\
\end{pmatrix}.
\end{align}
This means that the transformation matrix $\mathbf{Q}$ can, in principle, only be determined up to a unitary block-diagonal matrix.

In addition to the non-uniqueness of the block-diagonalization matrix $\mathbf{Q}$, there also exist multiple algorithms to determine such a matrix.
For example, one may subject the Fock matrix $\breve{\mathbf{F}}$ to subsequent Givens rotations \cite{Givens58}
to eliminate those off-diagonal matrix elements which represent the system-environment coupling. 
This is essentially equivalent to the Jacobi eigenvalue algorithm\cite{Jacobi46}, targeting only elements in the off-diagonal blocks.
However, this iterative procedure is tedious and may have difficulties to achieve off-diagonal blocks to be sufficiently close to zero.
Another example is the Householder block-diagonalization algorithm in which block reflectors are applied in an iterative fashion to eliminate off-diagonal blocks.\cite{Robbe05} 
However, it suffers from the same shortcomings as the Jacobi algorithm.
Also, all of these algorithms do not offer any kind of control and physical insight into how the subsystem and environment are separated from one another.

In an attempt to obtain a well-defined block-diagonalization matrix $\mathbf{Q}$ we require it to fulfill additional constraints.
For practical purposes it may be most convenient to have a transformed basis $\tilde{B}$ that is as close as possible to the initial basis $\breve{B}$.
This means that the block-diagonalization matrix $\mathbf{Q}$ is as close as possible to the identity matrix,
\begin{align}
\mathbf{Q} = \mathop{\mathrm{argmin}}_\mathbf{Q} \left( \left\| \mathbf{Q} - \mathbf{I} \right\| \right).
\label{eq:Q_cond}
\end{align}
Under these circumstances, matrices will be subject to minimal changes upon such a transformation, leaving their matrix structure mostly intact.
Subsequent approximations that are based on such a minimal-effect separation scheme can be expected to feature smallest errors.

In the following, we discuss how such a minimal-effect block-diagonalization matrix $\mathbf{Q}$ 
can be constructed for which the condition in Eq.~\eqref{eq:Q_cond} has been proven to be fulfilled by Cederbaum et al.\cite{Cederbaum89}

First, we express the block-diagonalization in Eq.~\eqref{eq:BD} in terms of the eigenvectors $\breve{\mathbf{C}}$ and $\tilde{\mathbf{C}}$,
\begin{align}
\tilde{\mathbf{F}} = 
\tilde{\mathbf{C}} \boldsymbol{\epsilon} \tilde{\mathbf{C}}^\mathsf{T} = 
\mathbf{Q} \breve{\mathbf{C}} \boldsymbol{\epsilon} \breve{\mathbf{C}}^\mathsf{T} \mathbf{Q}^\mathsf{T},
\end{align}
with $\tilde{\mathbf{C}} = \mathbf{Q} \breve{\mathbf{C}}$. Since the transformed Fock matrix $\tilde{\mathbf{F}}$ is block-diagonal, we may also write the eigenvector matrix $\tilde{\mathbf{C}}$ in a block-diagonal form,
\begin{align}
\tilde{\mathbf{C}} = \mathbf{Q} \breve{\mathbf{C}} = 
\begin{pmatrix}
\tilde{\mathbf{C}}_{\mathcal{S}} & \mathbf{0}\\
\mathbf{0} & \tilde{\mathbf{C}}_{\mathcal{E}}\\
\end{pmatrix}.
\end{align}
This means that the first $N_\mathcal{S}$ eigenvectors of $\breve{\mathbf{C}}$ are transformed into a basis $\tilde{B}$ in which they are located entirely on the transformed subsystem basis functions of $\tilde{B}_\mathcal{S}$. 
Accordingly, the last $N_\mathcal{E}$ eigenvectors of $\breve{\mathbf{C}}$ are transformed such that they are located entirely on the transformed environment basis functions of $\tilde{B}_\mathcal{E}$.
This assignment of eigenvectors implies that the eigenvector matrix $\breve{\mathbf{C}}$ may be written in terms of $\breve{\mathbf{C}}_\mathcal{S}$ and $\breve{\mathbf{C}}_\mathcal{E}$, such that
\begin{align}
\breve{\mathbf{C}} = 
\begin{pmatrix}
\breve{\mathbf{C}}_\mathcal{S} & \breve{\mathbf{C}}_\mathcal{E}
\end{pmatrix}. \label{eq:C_part}
\end{align} 
The partitioning of the basis $\tilde{B}$ into the subsets $\tilde{B}_\mathcal{S}$ and $\tilde{B}_\mathcal{E}$ is defined analogously to Eq.~\eqref{eq:basis_subsets},
\begin{align}
\tilde{B}_\mathcal{S} &= 
\left\lbrace\tilde{\phi}_i: i \in [1; N_\mathcal{S}]\right\rbrace,\\
\tilde{B}_\mathcal{E} &= 
\left\lbrace\tilde{\phi}_i: i \in [N_\mathcal{S}+1; N_B]\right\rbrace.
\end{align}
Rewriting the block-diagonalization of the Fock matrix in terms of the eigenvectors greatly simplifies the problem of determining the block-diagonalization matrix $\mathbf{Q}$. 
It also offers an additional layer of control considering the composition of subsystem and environment.
Since the order of eigenvectors in an eigenvector matrix is arbitrary, we can freely choose 
which eigenvectors of $\breve{\mathbf{C}}$ are projected into the subsystem or the environment basis.

To proceed with a unique construction of the block-diagonalization matrix $\mathbf{Q}$, we need to specify how eigenvectors from $\breve{\mathbf{C}}$ are assigned to either the subsystem $\breve{\mathbf{C}}_\mathcal{S}$ or the environment $\breve{\mathbf{C}}_\mathcal{E}$.
In this work, eigenvectors are assigned according to a simple localization scheme.
For each eigenvector $c_i$ there exists an associated localization function $f_i$ describing by
how much the first $N_\mathcal{S}$ basis functions of the orthogonalized basis contribute to the corresponding molecular orbital. The localization function employed in this work is given by,
\begin{align}
f_i = \sum_{j = 1}^{N_\mathcal{S}} \breve{c}_{i,j}^{2}.
\label{eq:cost_function}
\end{align}
It is, of course, possible to construct other localization functions that could be used to assign eigenvectors to either the subsystem or the environment.
Then, the eigenvector matrix $\breve{\mathbf{C}}$ is constructed as,
\begin{align}
\breve{\mathbf{C}} = 
\begin{pmatrix}
\breve{c}_1 & \breve{c}_2 & \cdots & \breve{c}_{N_B}
\end{pmatrix}.
\end{align}
where the eigenvectors are sorted in a descending order according to their localization function, such that
\begin{align}
f_{1} \geq f_{2} \geq \ldots \geq f_{N_B}.
\end{align}
This eigenvector ordering ensures that the $N_\mathcal{S}$ eigenvectors with the highest localization function are present in $\breve{\mathbf{C}}_\mathcal{S}$ and the remaining eigenvectors are present in $\breve{\mathbf{C}}_\mathcal{E}$.

Now that the eigenvector assignment has been discussed, we can continue with the construction of the unitary block-diagonalization matrix $\mathbf{Q}$. In the approach by Cederbaum et al.\ \cite{Cederbaum89}, it is constructed as a product of two matrices $\mathbf{Q}_\text{R}$ and $\mathbf{Q}_\text{BD}$,
\begin{align}
\mathbf{Q} = \mathbf{Q}_\text{R} \mathbf{Q}_\text{BD}.
\end{align}
This splits the procedure into two steps. First, matrix $\mathbf{Q}_\text{BD}$ block-diagonalizes the eigenvector matrix $\breve{\mathbf{C}}$. Then, the matrix $\mathbf{Q}_\text{R}$ renormalizes the transformation, guaranteeing that the total block-diagonalization matrix $\mathbf{Q}$ is unitary.
The matrix $\mathbf{Q}_\text{BD}$ takes on the form,
\begin{align}
\mathbf{Q}_\text{BD} =
\begin{pmatrix}
\mathbf{I} & -\mathbf{U}^\mathsf{T} \\
\mathbf{U} & \mathbf{I}
\end{pmatrix}.
\end{align}
It block-diagonalizes the eigenvector matrix $\breve{\mathbf{C}}$,
\begin{align}
\mathbf{Q}_\text{BD} \breve{\mathbf{C}} =
\begin{pmatrix}
- \breve{\mathbf{C}}_{11} + \mathbf{U}^\mathsf{T} \breve{\mathbf{C}}_{21} & - \breve{\mathbf{C}}_{12} + \mathbf{U}^\mathsf{T}  \breve{\mathbf{C}}_{22} \\
\mathbf{U} \breve{\mathbf{C}}_{11} + \breve{\mathbf{C}}_{21} & \mathbf{U} \breve{\mathbf{C}}_{12} + \breve{\mathbf{C}}_{22}
\end{pmatrix}.
\end{align}
Inspection of the off-diagonal blocks which are supposed to vanish, yields the following solution for $\mathbf{U}$,
\begin{align}
\mathbf{U} = 
- \breve{\mathbf{C}}_{21} \breve{\mathbf{C}}_{11}^{-1} = 
\left( \breve{\mathbf{C}}_{12} \breve{\mathbf{C}}_{22}^{-1} \right)^\mathsf{T}. 
\label{eq:cederbaumU}
\end{align}
Note that this matrix $\mathbf{U}$ can be constructed from either the subsystem eigenvectors $\breve{\mathbf{C}}_\mathcal{S}$ (as $\breve{\mathbf{C}}_{11}$ and $\breve{\mathbf{C}}_{21}$) or the environment eigenvectors $\breve{\mathbf{C}}_\mathcal{E}$ (as $\breve{\mathbf{C}}_{12}$ and $\breve{\mathbf{C}}_{22}$). A proof of this equality can be found in Appendix~\ref{app:u}.
The renormalization matrix $\mathbf{Q}_\text{R}$ is then given by,
\begin{align}
\mathbf{Q}_\text{R} 
&=
\left( \mathbf{Q}_\text{BD} \mathbf{Q}_\text{BD}^\mathsf{T} \right)^{-\frac{1}{2}} \label{eq:QR} \\
&=
\begin{pmatrix}
\left( \mathbf{I} + \mathbf{U}^\mathsf{T} \mathbf{U} \right)^{-\frac{1}{2}} & \mathbf{0} 
\\
\mathbf{0} & \left( \mathbf{I} + \mathbf{U} \mathbf{U}^\mathsf{T} \right)^{-\frac{1}{2}}
\end{pmatrix}
\end{align}
Finally, the block-diagonalization matrix $\mathbf{Q}$ reads,
\begin{align}
\mathbf{Q} =
\begin{pmatrix}
  (\mathbf{I} + \mathbf{U}^\mathsf{T} \mathbf{U})^{-\frac{1}{2}} & 
- (\mathbf{I} + \mathbf{U}^\mathsf{T} \mathbf{U})^{-\frac{1}{2}} \mathbf{U}^\mathsf{T} \\
  (\mathbf{I} + \mathbf{U} \mathbf{U}^\mathsf{T})^{-\frac{1}{2}} \mathbf{U} & 
  (\mathbf{I} + \mathbf{U} \mathbf{U}^\mathsf{T})^{-\frac{1}{2}}
\end{pmatrix}. \label{eq:Q}
\end{align}
In the literature concerning this kind of block-diagonalization,\cite{Cederbaum89,Sikkema09,Peng12,Seino12} there exist multiple notations for the actual form of the block-diagonalization matrix $\mathbf{Q}$. 
In Appendix~\ref{app:q} we show that all of these different representations are in fact identical.

Whereas L\"owdin's transformation is a minimal orthogonalization, the Cederbaum scheme represents a minimal transformation
in the sense that the new basis is as similar as possible to the old one. 
It is therefore no surprise that orbital localization schemes have been considered in the literature \cite{Zilkowski09,Li14}
that are based on a Cederbaum-type transformation, which relates to our approach here, but focuses on the preparation
of particular basis states rather than on the block-diagonalization of the Fock matrix.
Interestingly, Ref.\ \onlinecite{Li14} notes that the Cederbaum scheme can be
efficiently evaluated with techniques developed for the exact decoupling of the Dirac Hamiltonian, however without 
considering further implications on the relation of relativistic exact decoupling to a general embedding approach such as SSUB
introduced here.

\begin{figure*}[htbp!]
\includegraphics[width=\textwidth]{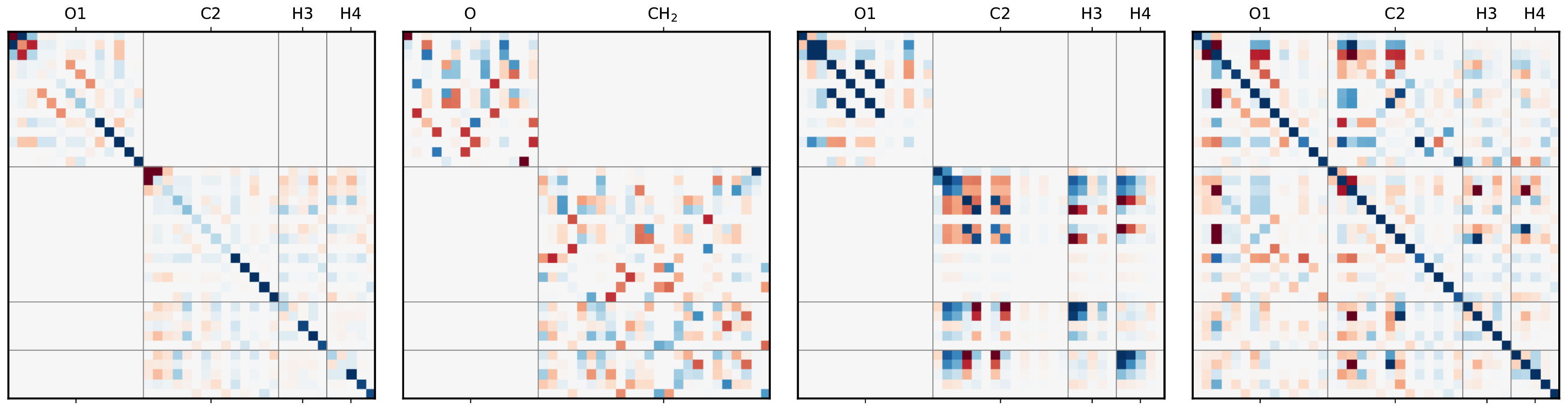}
\caption{The transformed Fock matrix $\tilde{\mathbf{F}}$, eigenvector matrix $\tilde{\mathbf{C}}$, density matrix $\tilde{\mathbf{P}}$, and transformation matrix $\mathbf{W}$ (from left to right) resulting from the matrices depicted in Fig.~\ref{fig:FCPS}. The block-diagonalization procedure was applied to decouple the oxygen atom from the other atoms of formaldehyde. Color coding as in Fig.~\ref{fig:FCPS} and for
$\mathbf{W}$ we have a range of $-$1 to +1.
}
\label{fig:FCPStrafo}
\end{figure*}

The total transformation matrix $\mathbf{W}$ transforming the original Fock matrix $\mathbf{F}$ into the block-diagonal form $\tilde{\mathbf{F}}$ then takes the form
\begin{align}
\mathbf{W} = \mathbf{Q} \mathbf{S}^{-\frac{1}{2}}.
\end{align}
Note that this transformation matrix $\mathbf{W}$ is, in general, not unitary.
The diagonal blocks $\tilde{\mathbf{F}}_\mathcal{S}$ and $\tilde{\mathbf{F}}_\mathcal{E}$ are given by
\begin{align}
\begin{split}
\tilde{\mathbf{F}}_\mathcal{S} =&
\mathbf{W}_{11} \mathbf{F}_{11} \mathbf{W}_{11}^\mathsf{T} + 
\mathbf{W}_{12} \mathbf{F}_{21} \mathbf{W}_{11}^\mathsf{T} \\
&+\mathbf{W}_{11} \mathbf{F}_{12} \mathbf{W}_{12}^\mathsf{T} + 
\mathbf{W}_{12} \mathbf{F}_{22} \mathbf{W}_{12}^\mathsf{T},
\end{split}
\label{eq:F_T_S}
\\
\begin{split}
\tilde{\mathbf{F}}_\mathcal{E} =&
\mathbf{W}_{21} \mathbf{F}_{11} \mathbf{W}_{21}^\mathsf{T} +
\mathbf{W}_{22} \mathbf{F}_{21} \mathbf{W}_{21}^\mathsf{T} \\
&+\mathbf{W}_{21} \mathbf{F}_{12} \mathbf{W}_{22}^\mathsf{T} + 
\mathbf{W}_{22} \mathbf{F}_{22} \mathbf{W}_{22}^\mathsf{T}.
\end{split}
\label{eq:F_T_E}
\end{align}\\
The one-electron matrix $\tilde{\mathbf{H}}$ and the two-electron matrix $\tilde{\mathbf{V}}$ are transformed in the same way as the Fock matrix $\tilde{\mathbf{F}}$. 
However, these matrices are in general not block-diagonal, only their sum $\tilde{\mathbf{F}}$ is.
Fig.~\ref{fig:FCPStrafo} provides a graphical representation of the transformed matrices from Fig.~\ref{fig:FCPS}.

We already note at this stage that the approach by Cederbaum et al.\ closely relates to relativistic exact two-component methods, which we discuss later in this work. 
We will also show how the system-environment separation according to this scheme can be generalized toward 
an arbitrary number of subsystems.
Because of these facts, we underline the general applicability of the unitary block-diagonalization approach by assigning the general term 'subsystem separation by unitary block-diagonalization' or SSUB (to be pronounced 'sub') to this approach.

\subsection{Partitioning into an arbitrary number of subsystems} \label{sec:multsub}
In the previous section, we considered a separation of a Fock matrix into two subsystems.
Starting from one of these subsystems, we may apply another unitary block-diagonalization step, which separates the subsystem into another two subsystems. 
Clearly, this allows one to split a total system into any number of subsystems. 
This process is only limited by the finite number of basis functions, which, in the limit, corresponds to the exact diagonalization of the Fock matrix, which has been the starting point for the construction of the unitary transformation matrices.

In the following, we discuss the technicalities of the separation of the system into multiple subsystems, following the formalism elaborated above.
The system shall be separated into $k$ subsystems, denoted by $\mathcal{S}_i$ each.
Each subsystem consists of $N_{\mathcal{S}_i}$ basis functions.
This generalization of the bipartition in section~\ref{sec:part} leads to the following 
partitioning of the basis,
\begin{align}
B_{\mathcal{S}_i} = \left\lbrace
\phi_i: i \in [1 + \sum_{j = 1}^{i-1} N_{\mathcal{S}_j}; \sum_{j = 1}^{i} N_{\mathcal{S}_j}]
\right\rbrace.
\end{align}

The eigenvectors in eigenvector matrix $\breve{\mathbf{C}}$ are localized such that they are located on the respective parts of the system.
Unfortunately, it is not trivial to assign each eigenvector to a single subsystem when partitioning the system into multiple subsystems.
A detailed description of how this was done in this work can be found in Appendix~\ref{app:localization}.
Analogously to the partitioning in Eq.~\eqref{eq:C_part}, this means that the eigenvector matrix can be written as,
\begin{align}
\breve{\mathbf{C}} = 
\begin{pmatrix}
\breve{\mathbf{C}}_{\mathcal{S}_1} & \breve{\mathbf{C}}_{\mathcal{S}_2} & \cdots & \breve{\mathbf{C}}_{\mathcal{S}_n}
\end{pmatrix},
\end{align}
where each of these eigenvector matrices $\breve{\mathbf{C}}_{\mathcal{S}_i}$ consists of $N_{\mathcal{S}_i}$ eigenvectors.

This ordered eigenvector matrix is then block-diagonalized into $k$ subsystems by a sequential application of a block-diagonalization matrix.
The total block-diagonalization matrix $\mathbf{Q}$ can be written as the product of each of these individual transformations,
\begin{align}
\mathbf{Q} = \overset{\curvearrowleft}{\prod^{k-1}_{i=1}} \mathbf{Q}^{(i)} = \mathbf{Q}^{(k-1)} \mathbf{Q}^{(k-2)} \cdots \mathbf{Q}^{(2)} \mathbf{Q}^{(1)}
\end{align}
Here, each block-diagonalization matrix $\mathbf{Q}^{(i)}$ produces the diagonal block $\tilde{\mathbf{C}}_{\mathcal{S}_i}$.
For the intermediary eigenvector matrices $\check{\mathbf{C}}^{(i)}$, we can write a recursion relation,
\begin{align}
\check{\mathbf{C}}^{(i + 1)} = \mathbf{Q}^{(i + 1)} \check{\mathbf{C}}^{(i)},
\end{align}
with $\check{\mathbf{C}}^{(0)} = \breve{\mathbf{C}}$ and $\check{\mathbf{C}}^{(k-1)} = \tilde{\mathbf{C}}$.
Each intermediary eigenvector matrix $\check{\mathbf{C}}^{(i)}$ can also be written as,
\begin{align}
\tilde{\mathbf{C}}^{(i)} =
\begin{pmatrix}
\tilde{\mathbf{C}}_{\mathcal{S}_1} &            & \mathbf{0}                         & \mathbf{0}                                   &        & \mathbf{0}                                 \\
                                   & \ddots     &                                    &                                              &        &                                            \\ 
\mathbf{0}                         &            & \tilde{\mathbf{C}}_{\mathcal{S}_i} & \mathbf{0}                                   &        & \mathbf{0}                                 \\
\mathbf{0}                         &            & \mathbf{0}                         & \check{\mathbf{C}}^{(i)}_{\mathcal{S}_{i+1}} & \cdots & \check{\mathbf{C}}^{(i)}_{\mathcal{S}_{k}}
\end{pmatrix}.
\end{align}

From this representation of the intermediary eigenvector matrices, we can construct the transformation matrix $\mathbf{Q}^{(i)}$ as,
\begin{align}
\mathbf{Q}^{(i)} =
\begin{pmatrix}
\mathbf{I}^{(i)} & \mathbf{0}            & \mathbf{0}            \\
\mathbf{0} & \mathbf{Q}^{(i)}_{11} & \mathbf{Q}^{(i)}_{12} \\
\mathbf{0} & \mathbf{Q}^{(i)}_{21} & \mathbf{Q}^{(i)}_{22}
\end{pmatrix},
\end{align}
with the matrix subblocks $\mathbf{Q}^{(i)}_{11}$, $\mathbf{Q}^{(i)}_{12}$, $\mathbf{Q}^{(i)}_{21}$ and $\mathbf{Q}^{(i)}_{22}$ being constructed exactly as in Eq.~\eqref{eq:Q}.
The dimension $d^{(i)}$ of the identity matrix subblock $\mathbf{I}^{(i)}$ is given by,
\begin{align}
d^{(i)} = \sum_{j=1}^{i-1} N_{\mathcal{S}_j}.
\end{align}
The matrix $\mathbf{U}^{(i)}$, which is required for the construction of the transformation matrix $\mathbf{Q}^{(i)}$ can be expressed (in analogy to Eq.~\eqref{eq:cederbaumU}) in terms of the intermediary subsystem eigenvector matrix $\check{\mathbf{C}}^{(i-1)}_{\mathcal{S}_{i}}$,
\begin{align}
\mathbf{U}^{(i)} = - 
\left( \check{\mathbf{C}}^{(i-1)}_{\mathcal{S}_{i}} \right)_\text{L}
\left( \check{\mathbf{C}}^{(i-1)}_{\mathcal{S}_{i}} \right)_\text{D}^{-1},
\end{align}
where the subscripts '$\text{D}$' and '$\text{L}$' denote the 
$N_{\mathcal{S}_i}{\times}N_{\mathcal{S}_i}$ diagonal block 
and the remaining lower off-diagonal block of the matrix $\check{\mathbf{C}}^{(i-1)}_{\mathcal{S}_{i}}$, respectively.

\begin{figure*}[htbp!]
\includegraphics[width=\textwidth]{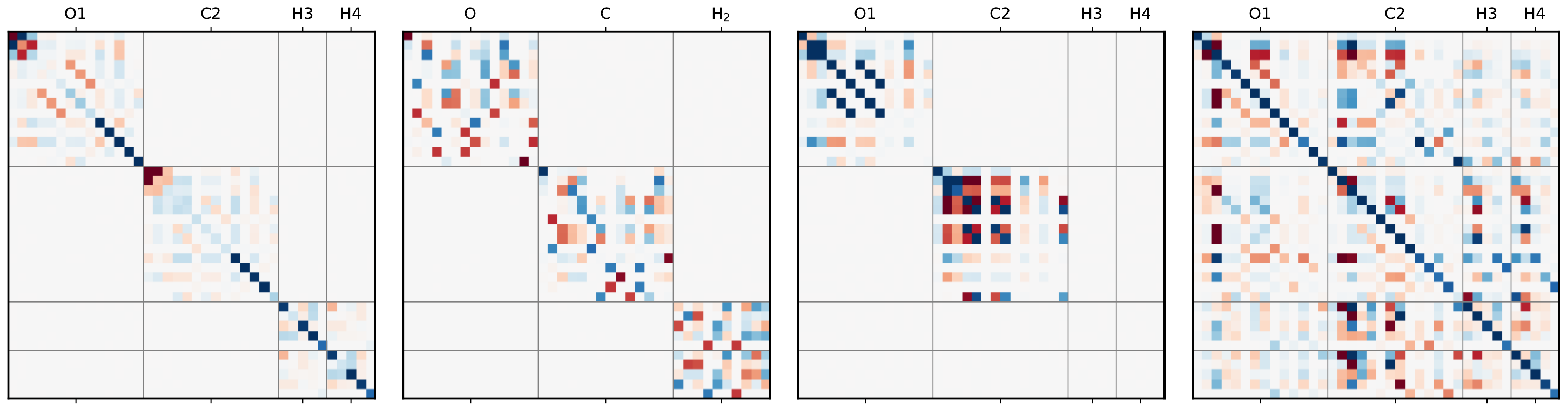}
\caption{The transformed Fock matrix $\tilde{\mathbf{F}}$, eigenvector matrix $\tilde{\mathbf{C}}$, density matrix $\tilde{\mathbf{P}}$, and transformation matrix $\mathbf{W}$ (from left to right) of formaldehyde, in which three subsystems were separated from each other: the oxygen atom, the carbon atom, and the hydrogen atoms.
Note that the hydrogen subblock of the density matrix $\tilde{\mathbf{P}}$ vanished.
This is because the eigenvectors which were assigned to this subsystem during the localization procedure are not occupied.
Instead, the occupied eigenvectors which contributed to the hydrogen atoms are now entirely located on the carbon-atom subsystem.
Note that this does not correspond to a charge transfer as it is simply a matter of representation in the transformed basis.
The color code is as in Fig. \ref{fig:FCPStrafo}.
\label{fig:3subsys}
}
\end{figure*}

\begin{figure*}[htbp!]
\includegraphics[width=\textwidth]{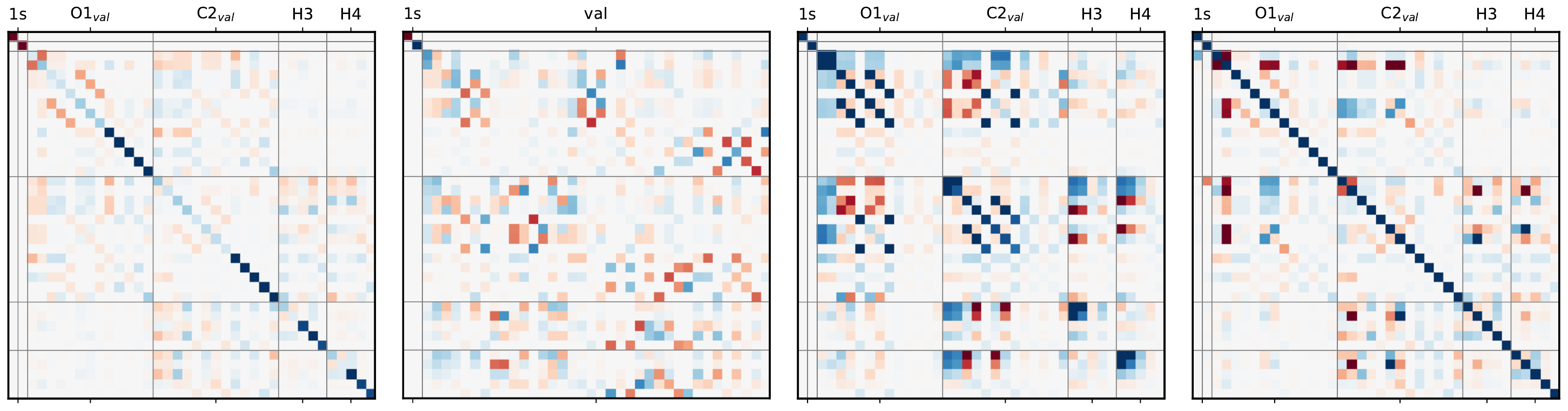}
\caption{
The transformed Fock matrix $\tilde{\mathbf{F}}$, eigenvector matrix $\tilde{\mathbf{C}}$, density matrix $\tilde{\mathbf{P}}$, and transformation matrix $\mathbf{W}$ (from left to right) of formaldehyde, in which the core eigenvectors of the carbon and oxygen atoms were 
separated from the rest of the system. This is an example of explicit core-valence separation.
The color code is as in Fig. \ref{fig:FCPStrafo}.
\label{fig:corevalence}
}
\end{figure*}

With this recursive scheme, it is now possible to construct a block-diagonalization matrix $\mathbf{Q}$ which block-diagonalizes the Fock matrix into $k$ subsystems.
\\
Two such sequential steps were carried out to arrive at a decomposition
of formaldehyde into three subsystems: the oxygen atom, the carbon atom, and the two hydrogen atoms (see Fig.~\ref{fig:3subsys}). 
As another example, we considered a core-valence separation. The subsystems consist 
of the eigenvectors that describe the two 1$s$-like molecular orbitals of the carbon and oxygen 
atoms and of all other orbitals (shown in Fig.~\ref{fig:corevalence}).

\section{Exploiting the block-diagonal structure in an approximate SCF procedure} \label{sec:BD-SCF}
We now proceed to introduce an approximate SCF scheme that avoids repeatedly solving the eigenvalue problem for the whole system.
The scheme is based on the block-diagonalization of the Fock matrix $\tilde{\mathbf{F}}$, leading to an advantageous representation of the eigenvalue problem encountered in the Roothaan--Hall equations.

With the initial density matrix $\mathbf{P}^{(0)}$ an initial Fock matrix $\mathbf{F}^{(1)}$ can be constructed.
Following the steps laid out in a previous section concerning block-diagonalization, we obtain the total transformation matrix $\mathbf{W} = \mathbf{Q} \mathbf{S}^{-\frac{1}{2}}$ from this Fock matrix $\mathbf{F}^{(1)}$.
In general, this step will be computationally expensive, since it involves (i) solving the 
eigenvalue problem for the whole system and (ii) 
diagonalizing the overlap matrix for the calculation of the inverse of its square-root required for $\mathbf{Q}$ in Eq.~\eqref{eq:Q}.
Clearly, computational advantages will only emerge for subsequent steps performed for a subsystem only rather than for the total system.

In principle, this SCF procedure follows the same steps as the SCF procedure with a transformed basis in section~\ref{sec:change_of_basis}.
However, the computational disadvantages arising from the additional forward and backward transformations in the exact formulation are alleviated by an advantageous representation of the eigenvalue problem, in which the dimension of the problem is significantly reduced:

\subsection{An approximate SCF procedure}
The transformed Fock matrix $\tilde{\mathbf{F}}^{(n)}$ is obtained by transforming the Fock matrix $\mathbf{F}^{(n)}$ with the transformation matrix $\mathbf{W}$.
Since the Fock matrix varies throughout the SCF procedure, exact block-diagonalization would require the construction of a block-diagonalization matrix $\mathbf{Q}$ in each iteration step.
This requires the solution of the full eigenvalue problem and subsequent matrix inversions and therefore the repeated evaluation of the block-diagonalization matrix $\mathbf{Q}$ would be computationally inefficient.
However, if we can construct a sufficiently good initial density matrix (such as one taken from a calculation on a very similar molecular structure
as, for instance, encountered in structure optimizations or first-principles molecular dynamics trajectories), 
we may expect changes in the Fock matrix over the course of the SCF procedure to be comparatively small. 
Then, the initial transformation matrix $\mathbf{W}$ can be kept over the course of a calculation and still block-diagonalize the resulting Fock matrices sufficiently well.
If the initial density matrix $\mathbf{P}^{(0)}$ deviates too much from the converged one, the proposed scheme will fail to find the self-consistent solution.

The diagonal blocks $\tilde{\mathbf{F}}^{(n)}_{\mathcal{S}}$ and $\tilde{\mathbf{F}}^{(n)}_{\mathcal{E}}$ can be evaluated as in Eqs.~\eqref{eq:F_T_S} and \eqref{eq:F_T_E}, respectively. However, if we assume the environment to change much less than the subsystem, 
we may leave $\tilde{\mathbf{F}}^{(n)}_{\mathcal{E}}$ constant,
\begin{align}
\tilde{\mathbf{F}}^{(n)}_{\mathcal{E}} = \tilde{\mathbf{F}}^{(1)}_{\mathcal{E}}.
\end{align}
Note that such an assumption may later be lifted in alternating freeze and thaw cycles.

Transforming the Roothaan--Hall equation with the transformation matrix $\mathbf{W} = \mathbf{Q} \mathbf{S}^{-\frac{1}{2}}$ according to Eq.~\eqref{eq:RHE_transform} requires us to solve an ordinary eigenvalue problem,
\begin{align}
\tilde{\mathbf{F}}^{(n)} \tilde{\mathbf{C}}^{(n)} &= \tilde{\mathbf{C}}^{(n)} \boldsymbol{\epsilon}^{(n)}.
\end{align}
Since the transformed Fock matrix $\tilde{\mathbf{F}}^{(n)}$ is block-diagonal, the whole eigenvalue problem assumes a block-diagonal form,
\begin{align}
\begin{split}
\begin{pmatrix}
\tilde{\mathbf{F}}^{(n)}_{\mathcal{S}} & \mathbf{0}\\
\mathbf{0} & \tilde{\mathbf{F}}^{(n)}_{\mathcal{E}}
\end{pmatrix} 
&\begin{pmatrix}
\tilde{\mathbf{C}}^{(n)}_{\mathcal{S}} & \mathbf{0}\\
\mathbf{0} & \tilde{\mathbf{C}}^{(n)}_{\mathcal{E}}
\end{pmatrix}
=\\
&\begin{pmatrix}
\tilde{\mathbf{C}}^{(n)}_{\mathcal{S}} & \mathbf{0}\\
\mathbf{0} & \tilde{\mathbf{C}}^{(n)}_{\mathcal{E}}
\end{pmatrix}
\begin{pmatrix}
\boldsymbol{\epsilon}^{(n)}_{\mathcal{S}} & \mathbf{0}\\
\mathbf{0} & \boldsymbol{\epsilon}^{(n)}_{\mathcal{E}}
\end{pmatrix}.
\end{split}
\end{align}
This allows us to separate the eigenvalue problem into two smaller eigenvalue problems
\begin{align}
\tilde{\mathbf{F}}^{(n)}_{\mathcal{S}} \tilde{\mathbf{C}}^{(n)}_{\mathcal{S}} = \tilde{\mathbf{C}}^{(n)}_{\mathcal{S}} \boldsymbol{\epsilon}^{(n)}_{\mathcal{S}} \label{eq:f_eig_sub}
\end{align}
and
\begin{align}
\tilde{\mathbf{F}}^{(n)}_{\mathcal{E}} \tilde{\mathbf{C}}^{(n)}_{\mathcal{E}} = \tilde{\mathbf{C}}^{(n)}_{\mathcal{E}} \boldsymbol{\epsilon}^{(n)}_{\mathcal{E}}, \label{eq:f_eig_env}
\end{align}
both of which can be solved separately.\\
However, if we set $\tilde{\mathbf{F}}^{(n)}_{\mathcal{E}} = \tilde{\mathbf{F}}^{(1)}_{\mathcal{E}}$, the second eigenvalue problem need not be solved as its solution has already been obtained for the construction of
$\mathbf{W}$.
The eigenvector matrix $\tilde{\mathbf{C}}^{(1)}_{\mathcal{E}}$ and the
eigenvalue matrix $\boldsymbol{\epsilon}^{(1)}_{\mathcal{E}}$ can be stored for later use.
Therefore, it is sufficient to solve the subsystem eigenvalue problem in Eq.~\eqref{eq:f_eig_sub}.

Selecting the occupied eigenvectors follows the standard SCF procedure. $N_\text{occ}$ eigenvectors with the lowest corresponding eigenvalues are occupied with electrons. 
As the eigenvalues and eigenvectors of the block-diagonal Fock matrix $\tilde{\mathbf{F}}^{(n)}$ can be assigned to either the subsystem or the environment,
it is possible to construct the occupied eigenvector matrices $\tilde{\mathbf{C}}^{(n)}_{\mathcal{S},\text{occ}}$ and $\tilde{\mathbf{C}}^{(n)}_{\mathcal{E},\text{occ}}$ from the eigenvector matrix blocks $\tilde{\mathbf{C}}^{(n)}_\mathcal{S}$ and $\tilde{\mathbf{C}}^{(n)}_\mathcal{E}$, respectively. 

The transformed density matrix $\tilde{\mathbf{P}}^{(n)}$ is calculated from the occupied eigenvectors according to Eq.~\eqref{eq:P_T_construction}. 
This matrix is block-diagonal since the eigenvector matrix $\tilde{\mathbf{C}}^{(n)}_\text{occ}$ takes a block-diagonal form. 
In terms of the occupied eigenvector matrices $\tilde{\mathbf{C}}^{(n)}_{\mathcal{S},\text{occ}}$ and $\tilde{\mathbf{C}}^{(n)}_{\mathcal{E},\text{occ}}$, 
$\tilde{\mathbf{P}}^{(n)}$ reads
\begin{align}
\hspace{0.5cm} 
\tilde{\mathbf{P}}^{(n)} = 2 \hspace{0.1cm} 
\begin{pmatrix}
\tilde{\mathbf{C}}^{(n)}_{\mathcal{S},\text{occ}} \left(\tilde{\mathbf{C}}^{(n)}_{\mathcal{S},\text{occ}}\right)^\mathsf{T} & \mathbf{0}\\
\mathbf{0} & \tilde{\mathbf{C}}^{(n)}_{\mathcal{E},\text{occ}} \left(\tilde{\mathbf{C}}^{(n)}_{\mathcal{E},\text{occ}}\right)^\mathsf{T}
\end{pmatrix}.
\end{align}

The density matrix $\mathbf{P}^{(n)}$ is obtained through back-transformation of the transformed density matrix $\tilde{\mathbf{P}}^{(n)}$.
Since the transformed density matrix $\tilde{\mathbf{P}}^{(n)}$ is block-diagonal, the back-transformation in Eq.~\eqref{eq:p_backtransform} can be simplified to yield the density matrix $\mathbf{P}^{(n)}$,
\begin{align}
\mathbf{P}^{(n)}
=
\begin{pmatrix}
\begin{matrix}
\mathbf{W}^\mathsf{T}_{11} \tilde{\mathbf{P}}^{(n)}_\mathcal{S} \mathbf{W}_{11} +\\ 
\mathbf{W}^\mathsf{T}_{21} \tilde{\mathbf{P}}^{(n)}_\mathcal{E} \mathbf{W}_{21} 
\end{matrix}
&
\hspace{0.1cm}
\begin{matrix}
\mathbf{W}^\mathsf{T}_{11} \tilde{\mathbf{P}}^{(n)}_\mathcal{S} \mathbf{W}_{12} +\\ 
\mathbf{W}^\mathsf{T}_{21} \tilde{\mathbf{P}}^{(n)}_\mathcal{E} \mathbf{W}_{22}
\end{matrix}
\vspace{0.25cm}
\\
\begin{matrix}
\mathbf{W}^\mathsf{T}_{12} \tilde{\mathbf{P}}^{(n)}_\mathcal{S} \mathbf{W}_{11} +\\ 
\mathbf{W}^\mathsf{T}_{22} \tilde{\mathbf{P}}^{(n)}_\mathcal{E} \mathbf{W}_{21}
\end{matrix}
&
\hspace{0.1cm}
\begin{matrix}
\mathbf{W}^\mathsf{T}_{12} \tilde{\mathbf{P}}^{(n)}_\mathcal{S} \mathbf{W}_{12} +\\ 
\mathbf{W}^\mathsf{T}_{22} \tilde{\mathbf{P}}^{(n)}_\mathcal{E} \mathbf{W}_{22}
\end{matrix}
\end{pmatrix}. \label{eq:P_bwd_block}
\end{align}
Note that it will be possible to precompute the terms arising from $\tilde{\mathbf{P}}^{(n)}_\mathcal{E}$, 
if this part of the transformed density matrix is kept constant throughout the SCF procedure
(for a frozen environment).

\begin{figure*}[htbp!] 
\includegraphics[width=\textwidth]{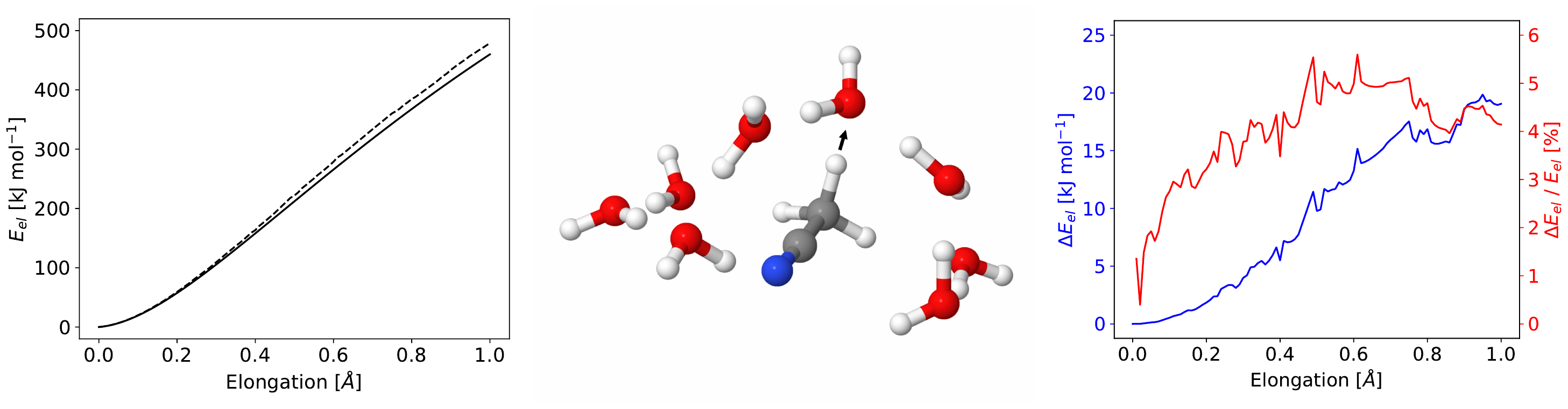}
\caption{Acetonitrile subsystem solvated in a water environment (middle). Electronic energy (given relative to that of the equilibrium structure) 
of the reference (solid) and approximate (dashed) SCF procedures for increasing C--H bond elongation (left). 
Error in electronic energy introduced by the approximate SCF scheme (right). \label{fig:perturbation}}
\end{figure*}

As an example, we chose a simple structure that represents a typical embedding situation, i.e., a solute surrounded by solvent molecules; here,
an acetonitrile molecule surrounded by seven water molecules. 
A  standard HF calculation is carried out on the initial equilibrium structure to obtain a converged density matrix. 
Then, one of the C--H bonds is elongated. 
For the subsequent structures with the elongated C--H bond, both a standard SCF calculation (as a reference for the exact electronic energy) 
and a calculation in the approximate SCF procedure exploiting the block-diagonal form of the transformed Fock matrix 
with a frozen environment of seven water molecules are performed. 
For the approximate calculation, the initial density matrix is taken from the converged calculation on the equilibrium structure performed initially. 
This allows us to demonstrate how large a structural perturbation may be in order for the approximate SCF to be sufficiently accurate.
The results of the calculations for the perturbed structures are summarized in Fig.~\ref{fig:perturbation}. 
Up to an elongation of around 0.3 \AA, the error is within chemical accuracy of 1 kcal mol$^{-1}$.

\subsection{Reducing the number of integrals in the two-electron matrix}\label{sec:V_approx}
The Fock matrix $\mathbf{F}^{(n)} = \mathbf{F}\farg{\mathbf{P}^{(n-1)}}$ is calculated from the density matrix $\mathbf{P}^{(n-1)}$.
When evaluating the two-electron matrix $\mathbf{V}^{(n)} = \mathbf{V}\farg{\mathbf{P}^{(n-1)}}$ it is possible to introduce approximations that make it possible to drastically reduce the number of required two-electron integrals. 

If subsystem and environment are well separated (either with respect to molecular structure in real space or in Hilbert space indicated by some energy gap), 
it will be possible to reduce the number of two-electron integrals required. 
This may facilitate the evaluation of the two-electron matrix $\mathbf{V}^{(n)} = \mathbf{V}\farg{\mathbf{P}^{(n-1)}}$ after the first SCF iteration step.

First, improvements in efficiency may be achieved by leaving the coupling and the pure-environment
blocks of the density matrix $\mathbf{P}$ unchanged,
\begin{align}
\mathbf{P}^{(n)} \approx
\begin{pmatrix}
\mathbf{P}_{11}^{(n)} & \mathbf{P}_{12}^{(0)} \\
\mathbf{P}_{21}^{(0)} & \mathbf{P}_{22}^{(0)}
\end{pmatrix}.
\end{align}
If $\mathbf{P}^{(n)}$ is approximated as shown above, it is not necessary to re-evaluate certain terms of the two-electron matrix in each SCF iteration step. For example,  
the Coulomb matrix $\mathbf{J}^{(n)}$ we may approximate as
\begin{align}
J_{ij}\farg{\mathbf{P}^{(n-1)}} \approx \sum^{N_\mathcal{S}}_{kl} P^{(n-1)}_{kl} \langle\phi_i\phi_j|\frac{1}{r_{12}}|\phi_k\phi_l\rangle\ + \delta J_{ij},
\end{align}
with the correction $\delta J_{ij}$ collecting all terms that are not purely subsystem-dependent
and may be evaluated only once in the beginning of the SCF procedure,
\begin{align}
\delta J_{ij} = J_{ij}\farg{\mathbf{P}^{(0)}} - \sum^{N_\mathcal{S}}_{kl} P^{(0)}_{kl} \langle\phi_i\phi_j|\frac{1}{r_{12}}|\phi_k\phi_l\rangle.
\end{align}
An analogous expression can be written for the exchange matrix $\mathbf{K}^{(n)}$.
For the exchange-correlation matrix $\mathbf{V}_\text{xc}$ in a KS-DFT framework, the approximation 
will take a slightly different form as only parts of the electron density do not need to be reevaluated,
\begin{align}
\rho\farg{\mathbf{P}^{(n-1)}} \approx \sum^{N_\mathcal{S}}_{kl} P^{(n-1)}_{kl} \phi_k \phi_l + \delta\rho,
\end{align}
with
\begin{align}
\delta\rho = \rho\farg{\mathbf{P}^{(0)}} - \sum^{N_\mathcal{S}}_{kl} P^{(0)}_{kl} \phi_k \phi_l.
\end{align}

In addition, the two-electron matrix $\mathbf{V}^{(n)}$ can be approximated by restricting the 
iterative re-evaluation to certain matrix elements 
of $\mathbf{V}^{(n)}$ only, most conveniently those in the $\mathbf{V}_{11}^{(n)}$ block,
\begin{align}
\mathbf{V}^{(n)} \approx
\begin{pmatrix}
\mathbf{V}_{11}^{(n)} & \mathbf{V}_{12}^{(1)} \\
\mathbf{V}_{21}^{(1)} & \mathbf{V}_{22}^{(1)}
\end{pmatrix},
\end{align}
which even further reduces the number of two-electron integrals to be evaluated.

\section{Projector-based embedding theory}\label{sec:MillerManby}
We are now in a position to discuss the embedding of a subsystem described 
by a high-level theory within the low-level mean-field method.
This allows us to directly relate our work to the recent work by Miller, Manby, and co-workers\cite{} as well as to Huzinaga and Cantu\cite{Huzinaga71}
and H\'egely et al.\cite{Hegely16}.
These embedding methods are based on an augmentation of the Fock operator with a projection operators. Therefore, we refer to them as projector-based embedding methods.

For projector-based embedding methods,
an initial calculation on the whole system with a low-level DFT method is usually carried out. 
Quantities from this initial calculation are denoted with a superscript $'(0)'$.
Then, the resulting eigenvectors in the eigenvector matrix $\mathbf{C}^{(0)}$ from this initial calculation are localized in such a way that they can be assigned to either the subsystem or the environment. 
This produces $\mathbf{C}^{(0)}_\mathcal{S}$ and $\mathbf{C}^{(0)}_\mathcal{E}$, which 
introduce a corresponding separation in the density matrix $\mathbf{P}^{(0)}$, being the 
direct sum of the two density matrices $\mathbf{P}^{(0)}_\mathcal{S}$ and $\mathbf{P}^{(0)}_\mathcal{E}$,
\begin{align}
\mathbf{P}^{(0)}_\mathcal{S} = 2 \hspace{0.1cm} \mathbf{C}^{(0)}_\mathcal{S,\text{occ}} \left( \mathbf{C}^{(0)}_\mathcal{S,\text{occ}} \right)^\mathsf{T}
\end{align}
and
\begin{align}
\mathbf{P}^{(0)}_\mathcal{E} = 2 \hspace{0.1cm} \mathbf{C}^{(0)}_\mathcal{E,\text{occ}} \left(\mathbf{C}^{(0)}_\mathcal{E,\text{occ}} \right)^\mathsf{T},
\end{align}
which implicitly assigns a number of electrons to the subsystem and the environment.

In the following, we seek the self-consistent solution of a composite Fock matrix $\mathbf{F}_\text{comp}$, mixing the two mean-field methods,
\begin{align}
\mathbf{F}_\text{comp} = \mathbf{F}^\text{Low}\farg{\mathbf{P}^{(0)}} + \mathbf{F}^\text{High}\farg{\mathbf{P}_\mathcal{S}} - \mathbf{F}^\text{Low}\farg{\mathbf{P}_\mathcal{S}^{(0)}}.
\end{align}
In projector-based embedding methods, it is required that the new subsystem eigenvectors in $\mathbf{C}_\mathcal{S}$ are orthogonal to the environment eigenvectors in $\mathbf{C}^{(0)}_\mathcal{E}$ (which are kept constant),
\begin{align}
\mathbf{C}_\mathcal{S}^\mathsf{T} \mathbf{S} \mathbf{C}^{(0)}_\mathcal{E} = \mathbf{0}. \label{eq:orthogonality}
\end{align}
The expression for the electronic energy of such a composite calculation is given by,
\begin{align}
E_\text{comp} = E^\text{Low}\farg{\mathbf{P}^{(0)}} + E^\text{High}\farg{\mathbf{P}_\mathcal{S}} - E^\text{Low}\farg{\mathbf{P}_\mathcal{S}^{(0)}}.
\end{align}

To ensure that orthogonality is preserved, the Fock matrix has to be modified even further such that its eigenvectors include the eigenvectors in $\mathbf{C}^{(0)}_\mathcal{E}$. This is done by augmenting the Fock operator with projection operators. In that case, solving the eigenvalue equation for this new Fock matrix $\mathbf{F}'$ will lead to subsystem eigenvectors $\mathbf{C}_\mathcal{S}$ that are necessarily orthogonal to those of the environment.
Such modified Fock matrices $\mathbf{F}'$ were proposed by Miller, Manby, and co-workers\cite{Manby12, Goodpaster14},
\begin{align}
\mathbf{F}'_\text{MM} = \mathbf{F}_\text{comp} + 
\mu \mathbf{S} \mathbf{C}^{(0)}_\mathcal{E} {\mathbf{C}^{(0)}_\mathcal{E}}^\mathsf{T} \mathbf{S},
\end{align}
and also by Huzinaga and Cantu\cite{Huzinaga71, Hegely16},
\begin{align}
\mathbf{F}'_\text{HC} = \mathbf{F}_\text{comp} - 
\mathbf{S} \mathbf{C}^{(0)}_\mathcal{E} {\mathbf{C}^{(0)}_\mathcal{E}}^\mathsf{T} \mathbf{F}_\text{comp} - 
\mathbf{F}_\text{comp} \mathbf{C}^{(0)}_\mathcal{E} {\mathbf{C}^{(0)}_\mathcal{E}}^\mathsf{T} \mathbf{S}.
\end{align}
Whereas the former is approximate, the latter is exact.
\\
The factor $\mu$ introduced in the Miller--Manby Fock matrix $\mathbf{F}'_\text{MM}$ can be understood as an energy shift which is applied to the molecular orbitals of the environment.
In principle, in the limit of $\mu$ tending to infinity, exact results can be obtained.
However, due to numerical issues, it is recommended to use values for $\mu$ of a few thousand Hartree.
\\
The Huzinaga--Cantu matrix $\mathbf{F}'_\text{HC}$ commutes with the matrix $\mathbf{S} \mathbf{C}^{(0)}_\mathcal{E} {\mathbf{C}^{(0)}_\mathcal{E}}^\mathsf{T} \mathbf{S}$ and therefore shares its eigenvectors $\mathbf{C}^{(0)}_\mathcal{E}$.

Here, we sketch how the additional term $\mu \mathbf{S} \mathbf{C}^{(0)}_\mathcal{E} {\mathbf{C}^{(0)}_\mathcal{E}}^\mathsf{T} \mathbf{S}$ in the Miller--Manby Fock matrix $\mathbf{F}'_\text{MM}$ forces the composite Fock matrix $\mathbf{F}_\text{comp}$ to assume the eigenvectors $\mathbf{C}^{(0)}_\mathcal{E}$.
We present an intuitive (but by no means rigorous) way to rationalize how these environment eigenvectors are recovered and how the energetic shift can be understood.

First, we realize that the generalized eigenvalue problem for $\mathbf{F}_\text{comp}$ in the Roothaan--Hall equation in Eq.~\eqref{eq:RHE} can also be written as,
\begin{align}
\mathbf{F}_\text{comp} = \mathbf{S} \mathbf{C} \boldsymbol{\epsilon} \mathbf{C}^\mathsf{T} \mathbf{S}.
\end{align}
Splitting this representation of $\mathbf{F}_\text{comp}$ into its subsystem and environment parts yields,
\begin{align}
\mathbf{F}_\text{comp} = 
\mathbf{S} \left(\mathbf{C}_\mathcal{S} \boldsymbol{\epsilon}_\mathcal{S} \mathbf{C}^\mathsf{T}_\mathcal{S} + \mathbf{C}_\mathcal{E} \boldsymbol{\epsilon}_\mathcal{E} \mathbf{C}^\mathsf{T}_\mathcal{E}\right) \mathbf{S}.
\end{align}
By adding the term $\mu \mathbf{S} \mathbf{C}^{(0)}_\mathcal{E} {\mathbf{C}^{(0)}_\mathcal{E}}^\mathsf{T} \mathbf{S}$, the modified Fock matrix $\mathbf{F}'_\text{MM}$ is obtained,
\begin{align}
\mathbf{F}'_\text{MM} = 
\mathbf{S} \left(
\mathbf{C}_\mathcal{S} \boldsymbol{\epsilon}_\mathcal{S} \mathbf{C}^\mathsf{T}_\mathcal{S} + \mathbf{C}_\mathcal{E} \boldsymbol{\epsilon}_\mathcal{E} \mathbf{C}^\mathsf{T}_\mathcal{E} +
\mu \mathbf{C}^{(0)}_\mathcal{E} {\mathbf{C}^{(0)}_\mathcal{E}}^\mathsf{T}
\right) \mathbf{S}.
\end{align}
Assuming that $\mathbf{C}^{(0)}_\mathcal{E} \approx \mathbf{C}_\mathcal{E}$ and that for all eigenvalues $|\epsilon_{ii}| \ll \mu$ holds, we arrive at 
\begin{align}
\mathbf{F}'_\text{MM} \approx 
\mathbf{S} \left(
\mathbf{C}_\mathcal{S} \boldsymbol{\epsilon}_\mathcal{S} \mathbf{C}^\mathsf{T}_\mathcal{S} + 
\mathbf{C}^{(0)}_\mathcal{E} (\boldsymbol{\epsilon}_\mathcal{E} + \mu \mathbf{I}) {\mathbf{C}^{(0)}_\mathcal{E}}^\mathsf{T}
\right) \mathbf{S}.
\end{align}
Since $\mu$ is such a large number (usually a few orders of magnitude larger than the largest eigenvalues), the eigenvectors $\mathbf{C}^{(0)}_\mathcal{E}$ dominate the matrix $\mathbf{F}'_\text{MM}$. In this representation, it is easy to see why $\mu$ is also referred to as an energy shift applied to the environment eigenvectors.

Interestingly, the projector-based embedding approach can be used in conjunction with our approximate SCF procedure. 
As noted in section~\ref{sec:BD-SCF}, in the approximate SCF procedure the same transformation matrix $\mathbf{W}$ 
may be applied throughout the whole procedure. Since the Fock matrix changes during the optimization, the block-diagonalization is not exact, introducing errors.
However, this is not the case for the modified Fock matrices $\mathbf{F}'_\text{MM}$ and $\mathbf{F}'_\text{HC}$. Since the environment eigenvectors $\mathbf{C}^{(0)}_\mathcal{E}$ are kept constant, exact block-diagonalization can be achieved for the whole SCF procedure. This implies that once the transformation matrix 
$\mathbf{W}$ has been calculated from the eigenvectors $\mathbf{C}^{(0)}_\mathcal{S}$ according to Eq.~\eqref{eq:Q}, projector-based embedding methods can be applied, solving the eigenvalue problem for the subsystem only.

We may now compare the Miller--Manby embedding method with the approximate block-diagonalized variant suggested in this section.
We chose again formaldehyde and partitioned it by cutting through the C=O double bond. 
This leaves us with a subsystem containing a single oxygen atom and the environment containing a carbon and two hydrogen atoms. 
The eigenvectors were localized according to the cost function in Eq.~\eqref{eq:cost_function}. 
For the high- and low-level mean-field schemes we simply chose Hartree--Fock and density functional theory with the BP86 functional\cite{Perdew86, Becke88}, respectively. 
We varied the parameter $\mu$ between 100 and 1000000 Hartree (values of at least a few thousand Hartree are recommended in Refs.\ \onlinecite{Manby12,Goodpaster14}).
The results are shown in Table~\ref{tab:MillerManby}. 
For values of $\mu$ above 10000 Hartree, it is possible to obtain the reference energy up to within five decimals with the approximate method.

\begin{table}[h]
\caption{Electronic energies obtained with the Miller--Manby projector based embedding method and their approximate subsystem SCF counterparts for different values for parameter $\mu$ (in Hartree) for a BP86-in-HF calculation of formaldehyde.\label{tab:MillerManby}}
\centering
\begin{tabular}{ p{.2\columnwidth} p{.3\columnwidth} p{.3\columnwidth}}
\hline
\hline
$\mu$ & MM & BD-MM\\
\hline 
100      & -113.7078797 & -113.7080317 \\ 
500      & -113.7106458 & -113.7106769 \\ 
1000     & -113.7109922 & -113.7110081 \\ 
5000     & -113.7112694 & -113.7112732 \\
10000    & -113.7113041 & -113.7113063 \\ 
100000   & -113.7113353 & -113.7113362 \\
1000000  & -113.7113384 & -113.7113391 \\
\hline
\hline
\end{tabular}
\end{table}

\section{Relation to embedded mean-field theory}
The embedded mean-field theory\cite{Fornace15} (EMFT) also aims at embedding a subsystem in an environment. 
EMFT is applied to describe the subsystem with a computationally expensive, high-level density functional within an environment described with a cheaper, lower-level density functional. 
Most importantly, in comparison to the projector-based embedding schemes, it is not necessary to calculate a self-consistent solution with the low-level method in advance.

\subsection{Embedded mean-field theory}\label{sec:EMFT}
The EMFT Fock matrix is a composite Fock matrix combining high- and low-level mean-field theories,
\begin{align}
\mathbf{F}^\text{EMFT}\farg{\mathbf{P}} = \mathbf{F}^\text{Low}\farg{\mathbf{P}} + \mathbf{F}^\text{High}_{11}\farg{\mathbf{P}_{11}} - \mathbf{F}^\text{Low}_{11}\farg{\mathbf{P}_{11}}.\label{eq:F_EMFT}
\end{align}
While this seems somewhat reminiscent of the Fock matrix expression in the projector-based embedding approach introduced earlier, it is different.
Here, $\mathbf{F}_{11}\farg{\mathbf{P}_{11}}$ denotes the $\mathbf{F}_{11}$ block of a Fock matrix where only the subsystem block of the density matrix was used to evaluate all contributions,
\begin{align}
F_{ij} = 
\begin{cases}
F_{ij}\left[ \left(
\begin{smallmatrix}
\mathbf{P}_{11} & \mathbf{0} \\
\mathbf{0} & \mathbf{0}
\end{smallmatrix}
\right) \right]
 &: i,j \leq N_\mathcal{S} \\
0 &: \rm else
\end{cases}.
\end{align}
This Fock matrix can be used in the SCF procedure as usual to obtain a self-consistent solution. In distinction to the projector-based embedding approaches in the previous section, no initial calculation on the whole system is required. Instead, both high- and low-level mean-field methods are converged to a self-consistent solution simultaneously.
The EMFT energy expression takes the composite form,
\begin{align}
E_\text{el}\farg{\mathbf{P}} = 
E^\text{Low}_\text{el}\farg{\mathbf{P}} +
E^\text{High}_\text{el}\farg{\mathbf{P}_{11}} - 
E^\text{Low}_\text{el}\farg{\mathbf{P}_{11}}. \label{eq:E_EMFT}
\end{align}

\subsection{EMFT with block-orthogonalized partitioning}
After the introduction of EMFT, it became apparent that some EMFT calculations exhibit an unphysical collapse of the self-consistent solution.\cite{Ding17} 
This has been attributed to a mismatch in the functional forms of the high- and low-level density functionals employed.\cite{Ding17}

The unphysical collapse manifests itself in a lowering of the electronic energy of several thousand Hartree.
Also, electron population analysis reveals that the collapse is accompanied by extraordinarily high electron populations in the subsystem and environment blocks, $\text{Tr}\left(\mathbf{P}_{11}\mathbf{S}_{11}\right)$ and $\text{Tr}\left(\mathbf{P}_{22}\mathbf{S}_{22}\right)$, respectively. 
Since the total number of electrons is constant, large negative populations in the coupling blocks $\text{Tr}\left(\mathbf{P}_{12}\mathbf{S}_{21}\right)$ and $\text{Tr}\left(\mathbf{P}_{21}\mathbf{S}_{12}\right)$ are generated. 
As a consequence, other observables are also affected and can be wrong by several orders of magnitude (consider, for example, the dipole moment).

To prevent this collapse of the self-consistent EMFT solution it was modified to operate on a different partitioning of the system.\cite{Ding17}
Instead of partitioning the system based on atomic orbitals, the scheme is applied in a block-orthogonalized basis (BOEMFT). 
This means that all matrix operators must be transformed accordingly.
The transformation matrix $\mathbf{W}$, which block-orthogonalizes the atomic orbital basis, is given by\cite{Ding17}
\begin{align}
\mathbf{W} = 
\begin{pmatrix}
\mathbf{I} & \mathbf{0}\\
- \mathbf{S}_{21} \mathbf{S}_{11}^{-1} & \mathbf{I}
\end{pmatrix}.
\end{align}
Transforming a basis $B$ into the basis $\tilde{B}$ with this transformation matrix $\mathbf{W}$ according to Eq.~\eqref{eq:basis_transformation} leaves the basis functions of the subsystem invariant.
Therefore, the subsystem subblock $\mathbf{A}_{11}$ is equal to the transformed subsystem block $\tilde{\mathbf{A}}_{11}$ of the transformed matrix $\tilde{\mathbf{A}}$.

Since the transformation is a block-orthogonalization, the transformed overlap matrix $\tilde{\mathbf{S}}$ is block-diagonal,
\begin{align}
\tilde{\mathbf{S}} =
\begin{pmatrix}
\mathbf{S}_{11} & \mathbf{0}\\
\mathbf{0} & \tilde{\mathbf{S}}_{22}
\end{pmatrix},
\end{align}
with
\begin{align}
\tilde{\mathbf{S}}_{22} = \mathbf{S}_{22} - \mathbf{S}_{21} \mathbf{S}_{11} \mathbf{S}_{12}.
\end{align}
In analogy to the EMFT Fock matrix in Eq.~\eqref{eq:F_EMFT}, the transformed BOEMFT Fock matrix $\tilde{\mathbf{F}}^\text{BOEMFT}$ is constructed according to
\begin{align}
\tilde{\mathbf{F}}^\text{BOEMFT}\farg{\mathbf{P}} = \tilde{\mathbf{F}}^\text{Low}\farg{\mathbf{P}} + \tilde{\mathbf{F}}^\text{High}_{11}\farg{\tilde{\mathbf{P}}_{11}} - \tilde{\mathbf{F}}^\text{Low}_{11}\farg{\tilde{\mathbf{P}}_{11}}.\label{eq:F_boemft}
\end{align}
Instead of the subblock $\mathbf{P}_{11}$ of the density matrix, the transformed density matrix subblock $\tilde{\mathbf{P}}_{11}$ is used. This is an approximation, 
producing a Fock matrix evaluated for unphysical subsystem electron densities.

The BOEMFT idea seems to resemble our approximate SCF procedure, with a basis transformation partitioning subsystem and environment. However, it is fundamentally different. In BOEMFT the transformation is only applied to prevent the unphysical collapse of the self-consistent field procedure,
whereas in our case it is the decisive starting point.

\section{Relation to relativistic decoupling theories}
The block-diagonalization of Dirac-based Fock operators has been a desire for physical, formal, and numerical
reasons as one wants to decouple electronic bound states from the negative-energy (continuum) states in order to arrive at variational 
electrons-only Hamiltonians suitable for applications in molecular physics and chemistry.\cite{Reiher09}

SSUB can be directly related to this block-diagonalization of the one-electron Dirac operator\cite{Heully86},
which is very efficiently formulated in basis-set representation.
In 2005, Jensen \cite{Jensen05}
proposed a scheme to decouple the negative-energy states (sometimes called 'positronic states') of the 
Dirac operator that avoids the involved algebra of Douglas--Kroll--Hess transformations.\cite{Hess86,wolf02b,reih04}
This work was driven by the desire of rewriting the Dirac Hamiltonian for electrons-only problems without introducing 
approximations \cite{bary97,bary01,bary02,fila03,fila03b,reih04,reih04b,reih06}.
Jensen started from a free-particle Foldy--Wouthuysen transformed Dirac Hamiltonian, which in Douglas--Kroll--Hess theory is the mandatory
first step \cite{reih04}. However, his central idea then was to 
construct the unitary matrix for the block-diagonalization of the four-component Dirac-based Fock operator
from its eigenvectors,
\begin{align}
\label{x2c}
\mathbf{U}^\text{X2C} = - \mathbf{C}_\text{SL} \mathbf{C}_\text{LL}^{-1},
\end{align}
(where $S$ and $L$ refer to the small- and large-components of
the molecular 4-spinors, respectively). 
It was then realized that the free-particle Foldy--Wouthuysen transformation can be skipped and Eq.\ (\ref{x2c})
can be applied directly to the Dirac-based Fock operator \cite{Kutzelnigg05,Ilias07,Saue11} (written in terms of Dyall's modified Dirac equation \cite{dyal97}).
This approach was later called exact two-component (X2C) decoupling.\cite{Kutzelnigg05,Liu06,Kutzelnigg06,Ilias07,Liu07,Sikkema09,Saue11,Peng12a}

As such, the X2C approach is identical to the block-diagonalization approach used by Cederbaum et al.\ \cite{Cederbaum89} earlier, but in a different context (cf.\ Eq.~\eqref{eq:cederbaumU}). 
Caution is advised when comparing different notations for the block-diagonalization matrices 
of Eq.~\eqref{eq:Q} with those encountered in the literature.\cite{Cederbaum89,Sikkema09,Peng12,Seino12}
At first glance, all of them seem to be slightly different, but they are, in fact, all equivalent.

X2C is usually viewed as a way to eliminate negative-energy states that are separated from the 'electronic' states by twice
the electron rest energy per electron in the system. This energy gap is huge because it depends, according to Einstein, on the
squared speed of light times the rest mass of an electron. The reduction of
4-spinors to 2-spinors essentially requires the elimination of all small-component basis functions. The contribution
of these basis functions to 'electronic', i.e., positive-energy molecular 4-spinors are usually atom centered and mostly
conserved in chemical reactions.\cite{Visscher97}
Accordingly, one is not surprised that, for physical reasons, X2C works so well at producing an effective electrons-only Hamiltonian
for 2-component calculations. However, in view of SSUB this separation of electronic and positronic states is nothing but a 
system-environment embedding, in which the effect of the positronic states is folded into the one-electron two-component
X2C Fock operator. Hence, no energy criteria need to be invoked to justify the separation of the negative-energy states, they are
only required to identify the one-particle eigenstates whose spinor energies are smaller than twice the
rest energy of an electron.

In fact, our analysis of SSUB shifts the focus from the physical picture to the actual mechanism in one-particle Hilbert space,
which allows us to better understand the decisive approximations that are in operation in practical X2C implementations.
SSUB applied to a four-component Dirac-based Fock operator produces the X2C Fock operator, but requires all eigenvectors of the
original four-component operator. This implies that the solution must already be known for the construction of
a unitarily transformed operator (note the relation to the formal projection operators built from the eigenstates
of the Dirac Hamiltonian proposed by Mittleman \cite{mitt72,mitt81} and
Sucher \cite{such80}; in particular, see Ref.\ \onlinecite{such87}). In relativistic quantum chemistry this
procedure is nevertheless advantageous as subsequent electron-correlation methods benefit largely from the elimination
of the negative-energy spinor states in the (preparatory) four-index transformation, which also holds true for general embedding schemes based
on SSUB-like ideas. 

\begin{figure*}[htbp!]
\includegraphics[width=\textwidth]{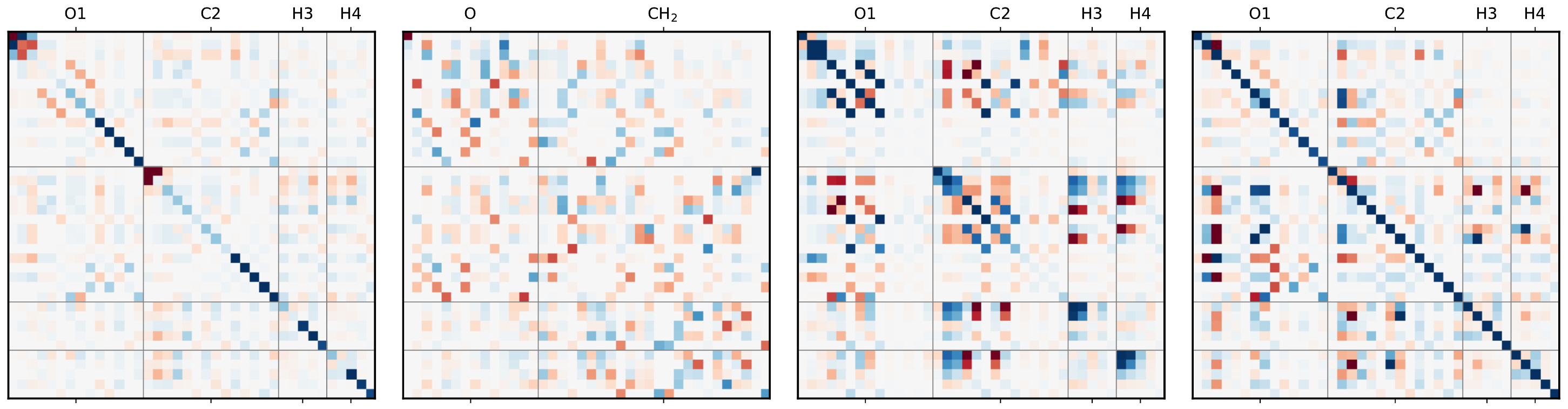}
\caption{
The transformed Fock matrix $\tilde{\mathbf{F}}$, eigenvector matrix $\tilde{\mathbf{C}}$, density matrix $\tilde{\mathbf{P}}$, and transformation matrix $\mathbf{W}$ (from left to right) of formaldehyde, obtained
by decoupling the oxygen atom from the other atoms with a transformation matrix $\mathbf{W}$ 
constructed from the eigenvectors of the orthogonalized one-electron matrix $\breve{\mathbf{H}}$,
which is an approximation inspired by standard decoupling methods considering only the nuclear potential for the block diagonalization of Dirac-based
Fock operators. 
A transformation based on the one-electron matrix $\mathbf{H}$ only is advantageous as 
it does not require a self-consistent solution. 
However, this approximate scheme does not yield block-diagonal matrices.
The color code is as in Fig. \ref{fig:FCPStrafo}.
\label{fig:coreapproxU}
}
\end{figure*}

Approximate approaches are usually applied that restrict and/or model the contributions to the potential. 
Various such approximations were proposed to obtain approximate eigenvectors for the construction of the unitary matrix in Eq.~\eqref{eq:Q}.
The most popular one, which is also applied in standard applications of sequential Douglas--Kroll--Hess decoupling transformations\cite{Hess86},
is the complete neglect of all electron-electron interaction terms, which alleviates the problem of obtaining a transformed
form of these terms. In other words, the iterative construction of the unitary matrix discussed in the context of the modified SCF scheme in section~\ref{sec:BD-SCF} is then avoided. 
The unitary transformation is then constructed from eigenvectors obtained by diagonalizing an approximate Roothaan--Hall
equation with the external electron-nuclei potential as the only potential.
Naturally, attempts were made to improve on this restricted model by introducing mean-field potentials which affect the eigenvectors
and, hence, the unitary block-diagonalizing matrix.\cite{hess96,schi98b,Sikkema09,Autschbach12}
As such, these approximate and popular versions of the X2C approach fall well into the class of approximate solutions that
we discussed in section~\ref{sec:BD-SCF} above.
In Fig.~\ref{fig:coreapproxU}, we demonstrate how relying only on the one-electron contributions of
Eq.~\eqref{eq:H_MF}, i.e., neglecting all other potential-energy contributions to the Fock operator,
affects the accuracy of the decoupling in general embedding schemes.
In comparison to the relativistic case, we see that this approximation does not work very well.
In the relativistic case, the eigenvectors are mainly dominated by the one-electron contributions
(because of the huge energy gap separating positive- and negative-energy states).
In mean-field theory, this is not the case.
The eigenvectors of the eigenvector matrix $\mathbf{H}$ and the Fock matrix $\mathbf{F}$ are fundamentally different.
For this reason, the approximation fails when applied to the general case, where no large energy gap 
sustains it.

\section{Approximations for huge single-particle spaces}
We have mentioned in the beginning that we will not address the issue of how to construct large Fock matrices that can then be subjected to SSUB.
However, given that one can construct very large Fock matrices (consider, for example, those in semi-empirical methods applied to molecular systems with
thousands of atoms), the issue of diagonalizing them will be pressing. Whereas one may apply subspace diagonalization to such large Fock matrices
(e.g., Lanczos or Davidson) and construct an approximate unitary matrix of reduced dimension, we may as well exploit approaches
developed to cope with such situations for X2C approaches.

The preparation of the X2C operator for large molecules has prompted the development of approximate decoupling schemes at the level of the Hamiltonian itself.\cite{Peralta04,Peralta05,Thar09}
A more accurate approximation is obtained at the level of the unitary transformation itself, which decomposes it into atomic blocks,\cite{Peng12,Seino12}
\begin{align}
\mathbf{Q} = \bigoplus_I \mathbf{Q}_{II} , \label{eq:DLU}
\end{align}
where $I$ denotes a subsystem (typically an atom) for which one-particle eigenstates have been determined.
which was called \textit{diagonal local approximation to the unitary decoupling transformation} (DLU) in Ref.\ \onlinecite{Peng12}
and \textit{local unitary transformation} (LUT) in Ref.\ \onlinecite{Seino12}.

\subsection{Local SSUB}
Here, we introduce the concept of an approximative local construction of the block-diagonalization matrix $\mathbf{Q}$, similar to the DLU and LUT schemes.
Given the fact that two names, DLU and LUT, are in use in relativistic quantum chemistry for the same idea and that this idea is important in the more general context of SSUB we may propose to call it L-SSUB for local approximation to SSUB.

\subsubsection{Fragmenting the Basis}
The local construction of the block-diagonalization matrix $\mathbf{Q}$ is based on the partitioning of the basis $B$ into $k$ fragment bases $B^{\mathcal{F}_i}$, such that
\begin{align}
B = \bigcup\limits_{i=1}^{k} B^{\mathcal{F}_i},
\end{align}
with
\begin{align}
B^{\mathcal{F}_i} \cap B^{\mathcal{F}_j} &= \varnothing.
\end{align}
Each fragment basis $B^{\mathcal{F}_i}$ consists of $N^{\mathcal{F}_i}$ basis functions.
It can be partitioned into $B^{\mathcal{F}_i}_\mathcal{S}$ and $B^{\mathcal{F}_i}_\mathcal{E}$, both of which are subsets of $B_\mathcal{S}$ and $B_\mathcal{E}$, respectively,
\begin{align}
B^{\mathcal{F}_i}_\mathcal{S} &= B^{\mathcal{F}_i} \cap B_\mathcal{S},\\
B^{\mathcal{F}_i}_\mathcal{E} &= B^{\mathcal{F}_i} \cap B_\mathcal{E}.
\end{align}
The fragment basis should be chosen such that it captures the dominant interactions between subsystem and environment.

\subsubsection{Constructing the local block-diagonalization matrix}
A fragment Fock matrix $\mathbf{F}^{\mathcal{F}_i}$ is constructed from a converged mean-field calculation employing the fragment basis $B^{\mathcal{F}_i}$ only.
For this mean-field calculation, a modified Fock operator may be applied, considering a subset of nuclei only, not taking those nuclei into account on which the basis functions are not located.
This fragment Fock matrix is then block-diagonalized with the block-diagonalization matrix $\mathbf{Q}^{\mathcal{F}_i}$ as outlined in section~\ref{sec:FBD}.
If a fragment basis $B^{\mathcal{F}_i}$ consists exclusively of subsystem or environment basis functions, the matrix $\mathbf{Q}^{\mathcal{F}_i}$ becomes the identity matrix.

With all fragment block-diagonalization matrices $\mathbf{Q}^{\mathcal{F}_i}$, the local block-diagonalization matrix $\mathbf{Q}^\mathcal{F}$ can be constructed in analogy to Eq.~\eqref{eq:DLU},
\begin{align}
\mathbf{Q}^\mathcal{F} = \bigoplus_{i=1}^{k} \mathbf{Q}^{\mathcal{F}_i}.
\end{align}
However, this local block-diagonalization matrix $\mathbf{Q}^\mathcal{F}$ cannot be applied to the Fock matrix $\mathbf{F}$ directly.

The matrix $\mathbf{Q}^\mathcal{F}$, is implicitly set up in a basis in which the order of basis functions is permuted.
Let $\hat{O}$ be an operator that reveals the order of a set in terms of its subsets.
While the basis order of the basis $B$, in which the Fock matrix $\mathbf{F}$ is represented, is given by
\begin{align}
\hat{O} B = \left[ B_\mathcal{S}, B_\mathcal{E} \right],
\end{align}
the fragmented basis $B^\mathcal{F}$ has the following order,
\begin{align}
\hat{O} B^\mathcal{F} = \left[ 
B^{\mathcal{F}_1}_\mathcal{S}, B^{\mathcal{F}_1}_\mathcal{E},
B^{\mathcal{F}_2}_\mathcal{S}, B^{\mathcal{F}_2}_\mathcal{E},
\cdots,
B^{\mathcal{F}_k}_\mathcal{S}, B^{\mathcal{F}_k}_\mathcal{E} 
\right].
\end{align}
While it seems to be complicated to convert the Fock matrix $\mathbf{F}$ to its fragment basis representation, it can actually be done with a rather trivial permutation.
Therefore, in order to block-diagonalize the Fock matrix $\mathbf{F}$ with the local block-diagonalization matrix $\mathbf{Q}^\mathcal{F}$, we first need to transform the Fock matrix $\mathbf{F}$ into the 
fragment basis representation with the permutation matrix $\mathbf{K}$. Afterwards, the block-diagonal Fock matrix $\tilde{\mathbf{F}}$ is recovered in the original basis $B$ by an inverse permutation,
\begin{align}
\tilde{\mathbf{F}} = \mathbf{K}^\mathsf{T} \mathbf{Q}^\mathcal{F} \mathbf{K} \mathbf{F} \mathbf{K}^\mathsf{T} \left(\mathbf{Q}^\mathcal{F}\right)^\mathsf{T} \mathbf{K}.
\end{align}
This implies that the total block-diagonalization matrix $\mathbf{Q}$ in the basis $B$ can be constructed as,
\begin{align}
\mathbf{Q} = \mathbf{K}^\mathsf{T} \mathbf{Q}^\mathcal{F} \mathbf{K},
\end{align}
which block-diagonalizes the Fock matrix directly.

\subsubsection{Quantifying approximate block-diagonalizations}
Since the local block-diagonalization is only approximate, we need to introduce some sort of measure that 
allows us to analyze the quality of the resulting block-diagonalization.
A naive approach to this is to simply inspect the off-diagonal blocks of the approximately block-diagonal 
Fock matrix $\tilde{\mathbf{F}}'$. 
However, such a measure would depend on the nature of the particular system since the magnitude of the matrix elements in the Fock matrix depend on the potentials in the Fock operator.
Here, we choose a diagonostic that has a more universal scope. 
It is based on a comparison between the eigenvectors of the approximately block-diagonalized Fock matrix $\tilde{\mathbf{F}}'$ and the exact block-diagonalized Fock matrix $\tilde{\mathbf{F}}$.
I.e., the measure $d_i$ quantifies the similarity of the eigenvectors $\tilde{c}_i$ and $\tilde{c}'_i$,
\begin{align}
d_i = \left| \tilde{c}_i^\mathsf{T} \tilde{c}'_i \right| ,
\end{align}
which is 1 for identical one-particle states.
To quantify the adequateness of a block-diagonalization for the whole system, we introduce the measure $D$, which is simply the average of all measures $d_i$ over all eigenvectors,
\begin{align}
D = N_B^{-1} \sum_{i=1}^{N_B} d_i.
\end{align}
The closer this measure $D$ is to $1$, the higher the accuracy of the approximate block-diagonalization.

\subsection{Example I: Structural separation}
\begin{figure*}[htbp!]
\includegraphics[width=\textwidth]{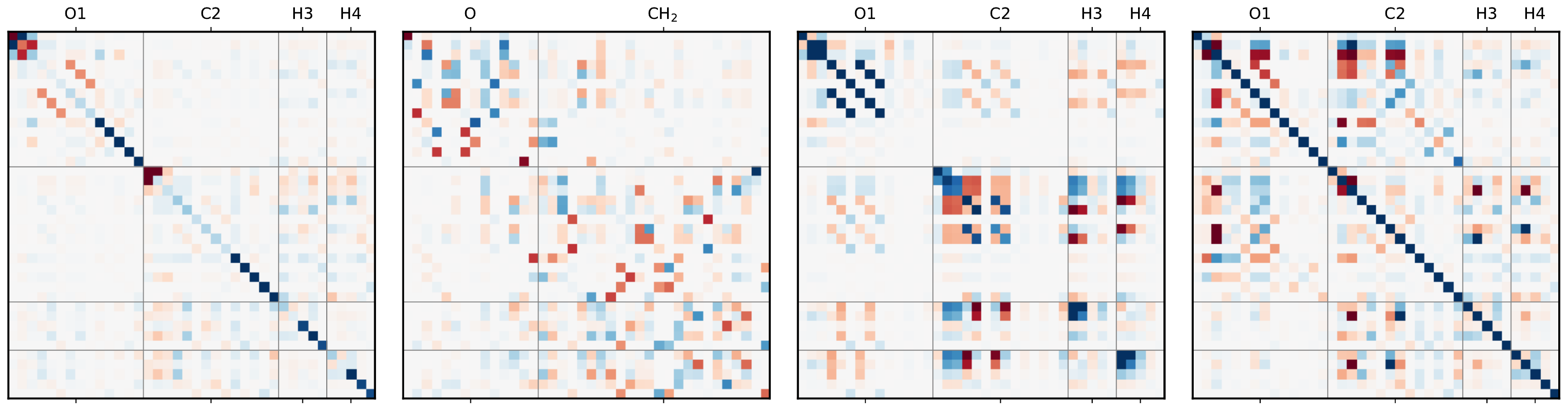}
\caption{
The transformed Fock matrix $\tilde{\mathbf{F}}$, eigenvector matrix $\tilde{\mathbf{C}}$, density matrix $\tilde{\mathbf{P}}$, and transformation matrix $\mathbf{W}$ (from left to right) of formaldehyde, obtained
by decoupling the oxygen atom from the other atoms with a block-diagonalization matrix $\mathbf{Q}$ obtained from a local approximation. The fragment on which the decoupling is based upon consists of only the oxygen and carbon atom. While the transformed Fock matrix $\tilde{\mathbf{F}}$ appears to be 
block-diagonal, the eigenvector matrix $\tilde{\mathbf{C}}$ and density matrix $\tilde{\mathbf{P}}$ have off-diagonal contributions.
The color code is as in Fig. \ref{fig:FCPStrafo}.
\label{fig:local_ssub}
}
\end{figure*}

Results of the local SSUB scheme for the formaldehyde molecule in which the oxygen atom is separated from the other atoms are shown in Fig.~\ref{fig:local_ssub}.
The fragment basis for which the block-diagonalization matrix was determined consists solely of the basis functions centered on the oxygen and the carbon atom. 

While this block-diagonalization is not exact, the decoupling scheme appears to decouple the most significant interactions and yields a Fock matrix $\tilde{\mathbf{F}}$ that for the most part is block-diagonal. 
However, the associated eigenvector matrix $\tilde{\mathbf{C}}$ and density matrix $\tilde{\mathbf{P}}$ show that there are interactions that could not be decoupled by this approach.
This observation is also supported by a relatively low measure $D~=~0.9222$. For certain eigenvectors, the measure $d$ is even below $0.7$, which indicates that the separation is not perfect.

\subsection{Example II: Core-Valence Separation}
\begin{figure*}[htbp!]
\includegraphics[width=\textwidth]{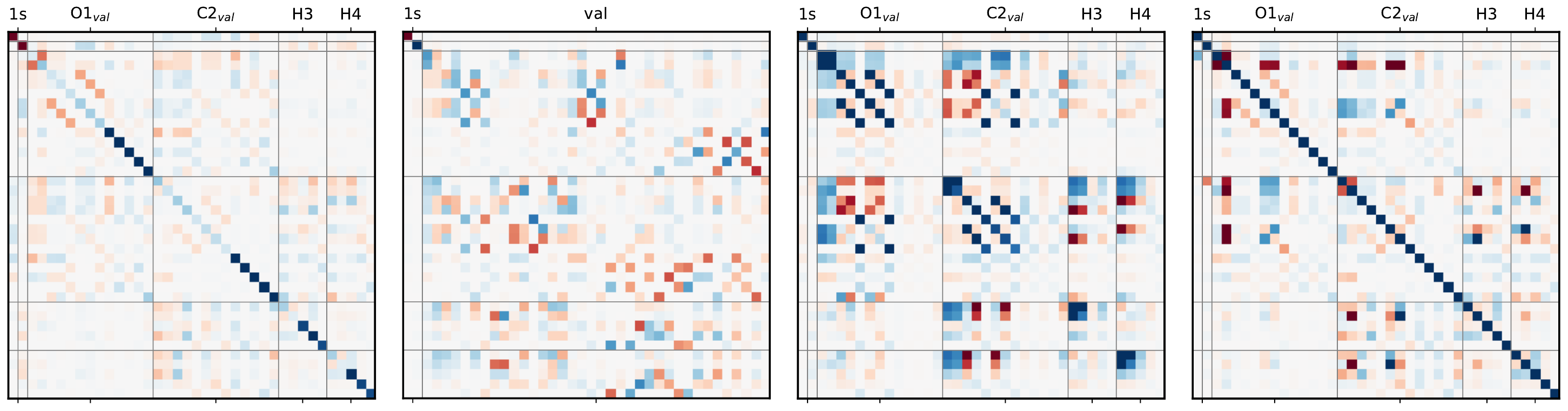}
\caption{
The transformed Fock matrix $\tilde{\mathbf{F}}$, eigenvector matrix $\tilde{\mathbf{C}}$, density matrix $\tilde{\mathbf{P}}$, and transformation matrix $\mathbf{W}$ (from left to right) of formaldehyde, obtained
by decoupling the 1$s$-like molecular orbitals of the carbon and oxygen atoms with a local approximation to the block-diagonalization matrix.
The color code is as in Fig. \ref{fig:FCPStrafo}.
\label{fig:core_valence_local}
}
\end{figure*}

Results of the local SSUB scheme for a core-valence separation applied to the formaldehyde molecule, in which the carbon and oxygen 1$s$-like molecular orbitals are separated from all other orbitals, are shown in Fig.~\ref{fig:core_valence_local}. 
The basis was split into three fragment bases. 
The first fragment basis consisted of the basis functions centered on the oxygen atom, the second one of the basis functions centered on the carbon atom.
The remaining hydrogen basis functions constitute the third fragment basis.
In this example, the transformed Fock matrix $\tilde{\mathbf{F}}$ shows a perturbed block-diagonal structure. 
However, the relatively high measure $D~=~0.9998$ implies that the separation was in fact successful.

\section{Conclusions}
In this work, we described a block-diagonalization approach directed toward an arbitrary
separation of a single-particle basis to define subsystems in a very general way, which we
call subsystem separation by unitary block-diagonalization (SSUB). The approach requires one
calculation on the full system and as such it shares this disadvantage with embedding
theories proposed by Manby and Miller and also with the relativistic exact two-component
approaches. However, the disadvantage is made up for by exploiting computational efficiency
for a subsystem in computational protocols that build upon a single-determinant solution.
For instance, these subsequent steps of a computational protocol may involve a four-index transformation
from the atomic orbital to a molecular orbital basis before a correlation calculation 
is considered. The latter is then significantly simplified for the subsystem (and usually would 
not have been possible for the total system, which is one reason for adopting an embedding approach).

The general nature of SSUB allows us to view the block-orthogonalization correction of EMFT from
a more general perspective. Moreover, projector-based embedding can be 
simplified significantly without compromising the inherent exactness of the method in the limit for large 
energy-shift parameter $\mu$. It also relates the field of embedding to very different ones
such as electron-positron separation in
relativistic quantum chemistry and purely electronic partitionings such as core-valence separation. 
As such, SSUB can be viewed as a framework for studying
quantum system partitioning at the single-particle level in a general way. 

We emphasize that a system separation by SSUB allows one to proceed with further calculations
on the separated parts of the system. In particular, it is possible to consider more accurate and, hence, computationally more expensive
ab initio calculations for a subsystem in the same spirit as in projector embedding and X2C theories.
Apart from this, one may exploit SSUB to accelerate standard protocols (e.g.,
for the solution of the SCF general eigenvalue problem).
In future work, we will therefore investigate a change of parameters for which the single-particle equations are formulated.
In particular, for accelerating calculations on a subsystem, a change of the nuclear framework --- as it occurs in structure optimizations,
first-principles molecular dynamics, and interactive reactivity studies --- represents such a change of parameters.
This will require a thorough study of the transferability of the system-separating matrix transformations in order to assess when
one may exploit a system separation, which has been obtained for some configuration, for related configurations.
Moreover, a generalization to time-dependent quantum dynamics should also be promising.


\begin{acknowledgments}
We gratefully acknowledge financial support by ETH Zurich (Grant Number ETH-46~16-2).
Part of this work was presented at the ICQC 2018 in Menton, France, in June 2018.
We thank Professor Trond Saue for remarks on the historical development of Jensen's original block-diagonalization
idea, mean-field spin--orbit Hamiltonians, and on two orbital localization schemes related to SSUB.
\end{acknowledgments}

\appendix
\section{Note concerning the construction of $\mathbf{U}$} \label{app:u}
Here, we show that the equality in Eq.~\eqref{eq:cederbaumU},
\begin{align}
\label{A1}
- \breve{\mathbf{C}}_{21} \breve{\mathbf{C}}_{11}^{-1} = \left( \breve{\mathbf{C}}_{12} \breve{\mathbf{C}}_{22}^{-1} \right)^\mathsf{T},
\end{align}
holds by considering the orthogonality of the eigenvector matrix $\breve{\mathbf{C}}$,
\begin{align}
\mathbf{I} = \breve{\mathbf{C}}^\mathsf{T} \breve{\mathbf{C}}.
\end{align}
This implies that subsystem and environment eigenvectors are orthogonal, leading to the expression,
\begin{align}
\mathbf{0} 
&= \breve{\mathbf{C}}_\mathcal{E}^\mathsf{T} \breve{\mathbf{C}}_\mathcal{S}\\
&= \breve{\mathbf{C}}_{12}^\mathsf{T} \breve{\mathbf{C}}_{11} + \breve{\mathbf{C}}_{22}^\mathsf{T} \breve{\mathbf{C}}_{21} \\
&= \breve{\mathbf{C}}_{22}^{-\mathsf{T}} \breve{\mathbf{C}}_{12}^\mathsf{T} + \breve{\mathbf{C}}_{21} \breve{\mathbf{C}}_{11}^{-1} \\
&= \left( \breve{\mathbf{C}}_{12} \breve{\mathbf{C}}_{22}^{-1} \right)^\mathsf{T} + \breve{\mathbf{C}}_{21} \breve{\mathbf{C}}_{11}^{-1},
\end{align}
from which Eq.~\eqref{A1} immediately follows.

\section{Representation of the block-diagonalization matrix $\mathbf{Q}$} \label{app:q}
In the literature,\cite{Cederbaum89,Sikkema09,Peng12,Seino12} there exist multiple representations of the block-diagonalization matrix $\mathbf{Q}$. 
These representations are very similar to the expression in Eq.~\eqref{eq:Q} and it is trivial to show that they achieve the exact same block-diagonalization.
The different representations arise from the following different definitions:
\begin{enumerate}
\item 
The definition of the block-diagonalization as either
\begin{align}
\tilde{\mathbf{F}} = \mathbf{Q} \mathbf{F} \mathbf{Q}^\mathsf{T},
\end{align}
or
\begin{align}
\tilde{\mathbf{F}} = \bar{\mathbf{Q}}^\mathsf{T} \mathbf{F} \bar{\mathbf{Q}},
\end{align}
with $\bar{\mathbf{Q}} = \mathbf{Q}^\mathsf{T}$.

\item 
The definition of the block-diagonalization matrix $\mathbf{Q}$ as either
\begin{align}
\mathbf{Q} = \mathbf{Q}_\text{R} \mathbf{Q}_\text{BD},
\end{align}
or
\begin{align}
\mathbf{Q} = \mathbf{Q}_\text{BD} \mathbf{Q}_\text{R}.
\end{align}
This is possible since $\mathbf{Q}_\text{R}$ and $\mathbf{Q}_\text{BD}$ commute. 

\item 
The definition of the block-diagonalization matrix $\mathbf{Q}_\text{BD}$ as either
\begin{align}
\mathbf{Q}_\text{BD} =
\begin{pmatrix}
\mathbf{I} & -\mathbf{U}^\mathsf{T} \\
\mathbf{U} & \mathbf{I}
\end{pmatrix},
\end{align}
or
\begin{align}
\mathbf{Q}_\text{BD} =
\begin{pmatrix}
\mathbf{I} & -\bar{\mathbf{U}} \\
\bar{\mathbf{U}}^\mathsf{T} & \mathbf{I}
\end{pmatrix},
\end{align}
with $\bar{\mathbf{U}} = \mathbf{U}^\mathsf{T}$. Since it is also possible to swap the sign of both $\bar{\mathbf{U}}$ and $\mathbf{U}$, this leaves us with four different definitions for the block-diagonalization matrix $\mathbf{Q}_\text{BD}$.
\end{enumerate}
All of these different definitions may be combined, which leads to the large number of possible representations of the block-diagonalization matrix $\mathbf{Q}$.

\section{Eigenvector assignment for multiple subsystems} \label{app:localization}
Assigning eigenvectors to multiple subsystems is not trivial when considering multiple subsystems.
When assigning an eigenvector to a subsystem with a localization function such as the one
in Eq.~\eqref{eq:cost_function}, we must consider that this eigenvector may also contribute to another subsystem.
Ranking these contributions may prove to be hard and can lead to the assignment of a single eigenvector to multiple subsystems. 
It is also possible that an eigenvector may not be chosen for any subsystem and is subsequently excluded from the block-diagonalization procedure.
This must not occur, 
and each eigenvector can be assigned to exactly one subsystem only.
Here, we introduce a scheme for an optimal and unique assignment of a set of eigenvectors to multiple subsystems.

We start by defining the localization function $f^{\mathcal{S}_i}_j$ for each eigenvector $\breve{c}_{j}$ for each subsystem $\mathcal{S}_i$,
\begin{align}
f^{\mathcal{S}_i}_j = \sum_{r = 1 + \sum_{t = 1}^{i-1} N_{\mathcal{S}_t}}^{\sum_{t = 1}^{i} N_{\mathcal{S}_t}} \breve{c}_{j,r}^{2}.
\end{align}
As in Eq.~\eqref{eq:cost_function}, this is simply a measure of the contribution of the subsystem basis functions to the corresponding molecular orbital.
After evaluation of the localization functions, the $N_{\mathcal{S}_i}$ eigenvectors with the highest localization function $f^{\mathcal{S}_i}_j$ are assigned to the subsystem $\mathcal{S}_i$.

However, this choice of eigenvectors suffers from the aforementioned problems concerning the assignment of eigenvectors to multiple subsystems.
This ambiguity can be resolved by the following iterative scheme:
\begin{enumerate}
\item For each pair of subsystems $\mathcal{S}_r$ and $\mathcal{S}_s$, the assignment of eigenvectors is probed for a duplicate assignment.
If no collision is detected, the procedure is completed. 
Otherwise, we continue with the next step.

\item One of the detected collisions is chosen at random. 
Let $\breve{c}_{i}$ be an eigenvector that was assigned to both subsystems $\mathcal{S}_r$ and $\mathcal{S}_s$,
but with a larger contribution to $\mathcal{S}_r$ than to $\mathcal{S}_s$,
\begin{align}
f^{\mathcal{S}_r}_i > f^{\mathcal{S}_s}_i.
\end{align}
In this case, the eigenvector $\breve{c}_{i}$ is removed from the choice of eigenvectors for subsystem $\mathcal{S}_s$.

\item Now that the subsystem $\mathcal{S}_s$ is missing an eigenvector, we need to assign a new eigenvector to this subsystem for it to be considered complete.
This new eigenvector $\breve{c}_{j}$ is chosen as the eigenvector with the highest localization function $f^{\mathcal{S}_s}_j$ which has not yet been chosen for subsystem $\mathcal{S}_s$.

\item The procedure is repeated from the first step to probe for new collisions since the new choice of eigenvectors is not guaranteed to be free of multiple assignments.
\end{enumerate}

Unless there is a case in which two localization functions for an eigenvector $\breve{c}_{i}$ are equal,
\begin{align}
f^{\mathcal{S}_r}_i = f^{\mathcal{S}_s}_i,
\end{align}
this scheme is guaranteed to produce a unique and correct assignment of the eigenvectors to multiple subsystems.
However, it should also be noted that in the case of an assignment of an eigenvector to multiple subsystems, caution is advised since it may be an indication of a particularly challenging partitioning.

\section*{References}

\begin{thebibliography}{100}

\bibitem{Cheng18}
{\sc C.~Cheng}, {\sc J.~Hussels}, {\sc M.~Niu}, {\sc H.~L. Bethlem}, {\sc
  K.~S.~E. Eikema}, {\sc E.~J. Salumbides}, {\sc W.~Ubachs}, {\sc M.~Beyer},
  {\sc N.~J. H{\"o}lsch}, {\sc J.~A. Agner}, {\sc F.~Merkt}, {\sc L.-G. Tao},
  {\sc S.-M. Hu}, and {\sc C.~Jungen},
\newblock {\em Phys. Rev. Lett.} {\bf 121}, 013001 (2018).

\bibitem{PhysRevA.97.060501}
{\sc L.~M. Wang} and {\sc Z.-C. Yan},
\newblock {\em Phys. Rev. A} {\bf 97}, 060501 (2018).

\bibitem{puch18}
{\sc M.~Puchalski}, {\sc A.~Spyszkiewicz}, {\sc J.~Komasa}, and {\sc
  K.~Pachucki},
\newblock {\em Phys. Rev. Lett.} {\bf 121}, 073001 (2018).

\bibitem{Barone15}
{\sc V.~Barone}, {\sc M.~Biczysko}, and {\sc C.~Puzzarini},
\newblock {\em Acc. Chem. Res.} {\bf 48}, 1413 (2015).

\bibitem{Puzzarini18}
{\sc C.~Puzzarini} and {\sc V.~Barone},
\newblock {\em Acc. Chem. Res.} {\bf 51}, 548 (2018).

\bibitem{Glowacki12}
{\sc D.~R. Glowacki}, {\sc C.-H. Liang}, {\sc C.~Morley}, {\sc M.~J. Pilling},
  and {\sc S.~H. Robertson},
\newblock {\em J. Phys. Chem. A} {\bf 116}, 9545 (2012).

\bibitem{Warshel76}
{\sc A.~Warshel} and {\sc M.~Levitt},
\newblock {\em J. Mol. Biol.} {\bf 103}, 227 (1976).

\bibitem{Singh86}
{\sc U.~C. Singh} and {\sc P.~A. Kollman},
\newblock {\em J. Comput. Chem.} {\bf 7}, 718 (1986).

\bibitem{Field90}
{\sc M.~J. Field}, {\sc P.~A. Bash}, and {\sc M.~Karplus},
\newblock {\em J. Comput. Chem.} {\bf 11}, 700 (1990).

\bibitem{Lin06}
{\sc H.~Lin} and {\sc D.~G. Truhlar},
\newblock {\em Theor. Chem. Acc.} {\bf 117}, 185 (2006).

\bibitem{Senn07b}
{\sc H.~M. Senn} and {\sc W.~Thiel},
\newblock {\em Curr. Opin. Chem. Biol.} {\bf 11}, 182 (2007).

\bibitem{Senn07a}
{\sc H.~M. Senn} and {\sc W.~Thiel},
\newblock {\em Top. Curr. Chem.} {\bf 268}, 173 (2007).

\bibitem{Senn09}
{\sc H.~M. Senn} and {\sc W.~Thiel},
\newblock {\em Angew. Chem. Int. Ed.} {\bf 48}, 1198 (2009).

\bibitem{Olsen10}
{\sc J.~M. Olsen}, {\sc K.~Aidas}, and {\sc J.~Kongsted},
\newblock {\em J. Chem. Theory Comput.} {\bf 6}, 3721 (2010).

\bibitem{Olsen11}
{\sc J.~M.~H. Olsen} and {\sc J.~Kongsted},
\newblock {\em Adv. Quantum Chem.} {\bf 61}, 107 (2011).

\bibitem{Sneskov11}
{\sc K.~Sneskov}, {\sc T.~Schwabe}, {\sc O.~Christiansen}, and {\sc
  J.~Kongsted},
\newblock {\em Phys. Chem. Chem. Phys.} {\bf 13}, 18551 (2011).

\bibitem{Svensson96}
{\sc M.~Svensson}, {\sc S.~Humbel}, {\sc R.~D.~J. Froese}, {\sc T.~Matsubara},
  {\sc S.~Sieber}, and {\sc K.~Morokuma},
\newblock {\em J. Phys. Chem.} {\bf 100}, 19357 (1996).

\bibitem{Dapprich99}
{\sc S.~Dapprich}, {\sc I.~Kom\'airomi}, {\sc K.~S. Byun}, {\sc K.~Morokuma},
  and {\sc M.~J. Frisch},
\newblock {\em J. Mol. Struct. THEOCHEM} {\bf 461-462}, 1 (1999).

\bibitem{Nakai02}
{\sc H.~Nakai},
\newblock {\em Chem. Phys. Lett.} {\bf 363}, 73 (2002).

\bibitem{Slipchenko06}
{\sc L.~V. Slipchenko} and {\sc M.~S. Gordon},
\newblock {\em J. Comput. Chem.} {\bf 28}, 276 (2006).

\bibitem{Gordon07}
{\sc M.~S. Gordon}, {\sc L.~Slipchenko}, {\sc H.~Li}, and {\sc J.~H. Jensen},
\newblock  {\bf 3}, 177 (2007).

\bibitem{Akama07}
{\sc T.~Akama}, {\sc M.~Kobayashi}, and {\sc H.~Nakai},
\newblock {\em J. Comput. Chem.} {\bf 28}, 2003 (2007).

\bibitem{Kobayashi09}
{\sc M.~Kobayashi} and {\sc H.~Nakai},
\newblock {\em Int. J. Quantum Chem.} {\bf 109}, 2227 (2009).

\bibitem{Gordon12}
{\sc M.~S. Gordon}, {\sc D.~G. Fedorov}, {\sc S.~R. Pruitt}, and {\sc L.~V.
  Slipchenko},
\newblock {\em Chem. Rev.} {\bf 112}, 632 (2012).

\bibitem{Breuer02}
{\sc H.-P. Breuer} and {\sc F.~Petruccione},
\newblock {\em The theory of open quantum systems},
\newblock Oxford University Press, 2002.

\bibitem{Amann11}
{\sc A.~Amann} and {\sc U.~M\"uller-Herold},
\newblock {\em Offene Quantensysteme},
\newblock Springer-Verlag Berlin Heidelberg, 2011.

\bibitem{White92}
{\sc S.~R. White},
\newblock {\em Phys. Rev. Lett.} {\bf 69}, 2863 (1992).

\bibitem{White93}
{\sc S.~R. White},
\newblock {\em Phys. Rev. B} {\bf 48}, 10345 (1993).

\bibitem{Schmidt07}
{\sc E.~Schmidt},
\newblock {\em Math. Ann.} {\bf 63}, 433 (1907).

\bibitem{Schollwoeck11}
{\sc U.~Schollw\"ock},
\newblock {\em Ann. Phys.} {\bf 326}, 96 (2011).

\bibitem{Knizia12}
{\sc G.~Knizia} and {\sc G.~K.-L. Chan},
\newblock {\em Phys. Rev. Lett.} {\bf 109}, 186404 (2012).

\bibitem{Knizia13}
{\sc G.~Knizia} and {\sc G.~K.-L. Chan},
\newblock {\em J. Chem. Theory Comput.} {\bf 9}, 1428 (2013).

\bibitem{Bulik14}
{\sc I.~W. Bulik}, {\sc G.~E. Scuseria}, and {\sc J.~Dukelsky},
\newblock {\em Phys. Rev. B} {\bf 89}, 035140 (2014).

\bibitem{Bulik14a}
{\sc I.~W. Bulik}, {\sc W.~Chen}, and {\sc G.~E. Scuseria},
\newblock {\em J. Chem. Phys.} {\bf 141}, 054113 (2014).

\bibitem{Welborn16}
{\sc M.~Welborn}, {\sc T.~Tsuchimochi}, and {\sc T.~Van~Voorhis},
\newblock {\em J. Chem. Phys.} {\bf 145}, 074102 (2016).

\bibitem{Ricke17}
{\sc N.~Ricke}, {\sc M.~Welborn}, {\sc H.-Z. Ye}, and {\sc T.~Van~Voorhis},
\newblock {\em Mol. Phys.} {\bf 115}, 2242 (2017).

\bibitem{Hohenberg64}
{\sc P.~Hohenberg} and {\sc W.~Kohn},
\newblock {\em Phys. Rev.} {\bf 136}, B864 (1964).

\bibitem{Kohn65}
{\sc W.~Kohn} and {\sc L.~J. Sham},
\newblock {\em Phys. Rev.} {\bf 140}, A1133 (1965).

\bibitem{Gordon72}
{\sc R.~G. Gordon} and {\sc Y.~S. Kim},
\newblock {\em J. Chem. Phys.} {\bf 56}, 3122 (1972).

\bibitem{Kim74}
{\sc Y.~S. Kim} and {\sc R.~G. Gordon},
\newblock {\em J. Chem. Phys.} {\bf 60}, 1842 (1974).

\bibitem{Senatore86}
{\sc G.~Senatore} and {\sc K.~Subbaswamy},
\newblock {\em Phys. Rev. B} {\bf 34}, 5754 (1986).

\bibitem{Cortona91}
{\sc P.~Cortona},
\newblock {\em Phys. Rev. B} {\bf 44}, 8454 (1991).

\bibitem{Wesolowski93}
{\sc T.~A. Wesolowski} and {\sc A.~Warshel},
\newblock {\em J. Phys. Chem.} {\bf 97}, 8050 (1993).

\bibitem{Cortona94}
{\sc P.~Cortona} and {\sc A.~V. Monteleone},
\newblock {\em Int. J. Quantum Chem.} {\bf 52}, 987 (1994).

\bibitem{Neugebauer05}
{\sc J.~Neugebauer}, {\sc M.~J. Louwerse}, {\sc E.~J. Baerends}, and {\sc T.~A.
  Wesolowski},
\newblock {\em J. Chem. Phys.} {\bf 122}, 094115 (2005).

\bibitem{Iannuzzi06}
{\sc M.~Iannuzzi}, {\sc B.~Kirchner}, and {\sc J.~Hutter},
\newblock {\em Chem. Phys. Lett.} {\bf 421}, 16 (2006).

\bibitem{Jacob08}
{\sc C.~R. Jacob}, {\sc J.~Neugebauer}, and {\sc L.~Visscher},
\newblock {\em J. Comput. Chem.} {\bf 29}, 1011 (2008).

\bibitem{Fux10}
{\sc S.~Fux}, {\sc C.~R. Jacob}, {\sc J.~Neugebauer}, {\sc L.~Visscher}, and
  {\sc M.~Reiher},
\newblock {\em J. Chem. Phys.} {\bf 132}, 164101 (2010).

\bibitem{Elliott10}
{\sc P.~Elliott}, {\sc K.~Burke}, {\sc M.~H. Cohen}, and {\sc A.~Wasserman},
\newblock {\em Phys. Rev. A} {\bf 82}, 024501 (2010).

\bibitem{Goodpaster10}
{\sc J.~D. Goodpaster}, {\sc N.~Ananth}, {\sc F.~R. Manby}, and {\sc T.~F.
  Miller},
\newblock {\em J. Chem. Phys.} {\bf 133}, 084103 (2010).

\bibitem{Jacob14}
{\sc C.~R. Jacob} and {\sc J.~Neugebauer},
\newblock {\em WIREs Comput Mol Sci} {\bf 4}, 325 (2014).

\bibitem{Fornace15}
{\sc M.~E. Fornace}, {\sc J.~Lee}, {\sc K.~Miyamoto}, {\sc F.~R. Manby}, and
  {\sc T.~F. Miller},
\newblock {\em J. Chem. Theory Comput.} {\bf 11}, 568 (2015).

\bibitem{Wesolowski15}
{\sc T.~A. Wesolowski}, {\sc S.~Shedge}, and {\sc X.~Zhou},
\newblock {\em Chem. Rev.} {\bf 115}, 5891 (2015).

\bibitem{Ding17}
{\sc F.~Ding}, {\sc F.~R. Manby}, and {\sc T.~F. Miller},
\newblock {\em J. Chem. Theory Comput.} {\bf 13}, 1605 (2017).

\bibitem{Huzinaga71}
{\sc S.~Huzinaga} and {\sc A.~A. Cantu},
\newblock {\em J. Chem. Phys.} {\bf 55}, 5543 (1971).

\bibitem{Govind99}
{\sc N.~Govind}, {\sc Y.~A. Wang}, and {\sc E.~A. Carter},
\newblock {\em J. Chem. Phys.} {\bf 110}, 7677 (1999).

\bibitem{Kluener02}
{\sc T.~Kl\"uner}, {\sc N.~Govind}, {\sc Y.~A. Wang}, and {\sc E.~A. Carter},
\newblock {\em J. Chem. Phys.} {\bf 116}, 42 (2002).

\bibitem{Huang06}
{\sc P.~Huang} and {\sc E.~A. Carter},
\newblock {\em J. Chem. Phys.} {\bf 125}, 084102 (2006).

\bibitem{Gomes08}
{\sc A.~S.~P. Gomes}, {\sc C.~R. Jacob}, and {\sc L.~Visscher},
\newblock {\em Phys. Chem. Chem. Phys.} {\bf 10}, 5353 (2008).

\bibitem{Huang11}
{\sc C.~Huang}, {\sc M.~Pavone}, and {\sc E.~A. Carter},
\newblock {\em J. Chem. Phys.} {\bf 134}, 154110 (2011).

\bibitem{Manby12}
{\sc F.~R. Manby}, {\sc M.~Stella}, {\sc J.~D. Goodpaster}, and {\sc T.~F.
  Miller},
\newblock {\em J. Chem. Theory Comput.} {\bf 8}, 2564 (2012).

\bibitem{Hoefener12}
{\sc S.~H\"ofener} and {\sc L.~Visscher},
\newblock {\em J. Chem. Phys.} {\bf 137}, 204120 (2012).

\bibitem{Goodpaster14}
{\sc J.~D. Goodpaster}, {\sc T.~A. Barnes}, {\sc F.~R. Manby}, and {\sc T.~F.
  Miller~III},
\newblock {\em J. Chem. Phys.} {\bf 140}, 18A507 (2014).

\bibitem{Daday14}
{\sc C.~Daday}, {\sc C.~K{\"o}nig}, {\sc J.~Neugebauer}, and {\sc C.~Filippi},
\newblock {\em ChemPhysChem} {\bf 15}, 3205 (2014).

\bibitem{Dresselhaus15}
{\sc T.~Dresselhaus}, {\sc J.~Neugebauer}, {\sc S.~Knecht}, {\sc S.~Keller},
  {\sc Y.~Ma}, and {\sc M.~Reiher},
\newblock {\em J. Chem. Phys.} {\bf 142}, 044111 (2015).

\bibitem{Hegely16}
{\sc B.~H\'egely}, {\sc P.~R. Nagy}, {\sc G.~G. Ferenczy}, and {\sc
  M.~K\'allay},
\newblock {\em J. Chem. Phys.} {\bf 145}, 064107 (2016).

\bibitem{Kananenka15}
{\sc A.~A. Kananenka}, {\sc E.~Gull}, and {\sc D.~Zgid},
\newblock {\em Phys. Rev. B} {\bf 91}, 121111 (2015).

\bibitem{Lan15}
{\sc T.~N. Lan}, {\sc A.~A. Kananenka}, and {\sc D.~Zgid},
\newblock {\em J. Chem. Phys.} {\bf 143}, 241102 (2015).

\bibitem{Lan17}
{\sc T.~N. Lan} and {\sc D.~Zgid},
\newblock {\em J. Phys. Chem. Lett.} {\bf 8}, 2200 (2017).

\bibitem{Georges92}
{\sc A.~Georges} and {\sc G.~Kotliar},
\newblock {\em Phys. Rev. B} {\bf 45}, 6479 (1992).

\bibitem{Georges96}
{\sc A.~Georges}, {\sc G.~Kotliar}, {\sc W.~Krauth}, and {\sc M.~J. Rozenberg},
\newblock {\em Rev. Mod. Phys.} {\bf 68}, 13 (1996).

\bibitem{Georges04}
{\sc A.~Georges},
\newblock {\em AIP Conf. Proc} {\bf 715}, 3 (2004).

\bibitem{Kotliar06}
{\sc G.~Kotliar}, {\sc S.~Y. Savrasov}, {\sc K.~Haule}, {\sc V.~S. Oudovenko},
  {\sc O.~Parcollet}, and {\sc C.~A. Marianetti},
\newblock {\em Rev. Mod. Phys.} {\bf 78}, 865 (2006).

\bibitem{Fromager15}
{\sc E.~Fromager},
\newblock {\em Mol. Phys.} {\bf 113}, 419 (2015).

\bibitem{Senjean17}
{\sc B.~Senjean}, {\sc M.~Tsuchiizu}, {\sc V.~Robert}, and {\sc E.~Fromager},
\newblock {\em Mol. Phys.} {\bf 115}, 48 (2017).

\bibitem{Senjean18}
{\sc B.~Senjean}, {\sc N.~Nakatani}, {\sc M.~Tsuchiizu}, and {\sc E.~Fromager},
\newblock {\em Phys. Rev. B} {\bf 97}, 235105 (2018).

\bibitem{Roos80}
{\sc B.~O. Roos}, {\sc P.~R. Taylor}, and {\sc P.~E.~M. Siegbahn},
\newblock {\em Chem. Phys.} {\bf 48}, 157 (1980).

\bibitem{Roos80a}
{\sc B.~O. Roos},
\newblock {\em Int. J. Quantum Chem.} {\bf 18}, 175 (1980).

\bibitem{Ruedenberg82}
{\sc K.~Ruedenberg}, {\sc M.~W. Schmidt}, {\sc M.~M. Gilbert}, and {\sc S.~T.
  Elbert},
\newblock {\em Chem. Phys.} {\bf 71}, 41 (1982).

\bibitem{Olsen88}
{\sc J.~Olsen}, {\sc B.~O. Roos}, {\sc P.~J{\o}rgensen}, and {\sc H.~J.~A.
  Jensen},
\newblock {\em J. Chem. Phys.} {\bf 89}, 2185 (1988).

\bibitem{Fleig03}
{\sc T.~Fleig}, {\sc J.~Olsen}, and {\sc L.~Visscher},
\newblock {\em J. Chem. Phys.} {\bf 119}, 2963 (2003).

\bibitem{Ivanic03}
{\sc J.~Ivanic},
\newblock {\em J. Chem. Phys.} {\bf 119}, 9364 (2003).

\bibitem{Parker13}
{\sc S.~M. Parker}, {\sc T.~Seideman}, {\sc M.~A. Ratner}, and {\sc
  T.~Shiozaki},
\newblock {\em J. Chem. Phys.} {\bf 139}, 021108 (2013).

\bibitem{Parker14}
{\sc S.~M. Parker} and {\sc T.~Shiozaki},
\newblock {\em J. Chem. Theory Comput.} {\bf 10}, 3738 (2014).

\bibitem{Parker14a}
{\sc S.~M. Parker} and {\sc T.~Shiozaki},
\newblock {\em J. Chem. Phys.} {\bf 141}, 211102 (2014).

\bibitem{Thiel88}
{\sc W.~Thiel},
\newblock {\em Tetrahedron} {\bf 44}, 7393 (1988).

\bibitem{Dewar92}
{\sc M.~J.~S. Dewar},
\newblock {\em Int. J. Quantum Chem.} {\bf 44}, 427 (1992).

\bibitem{Clark93}
{\sc T.~Clark},
\newblock {\em Semiempirical Molecular Orbital Theory: Facts, Myths and
  Legends}, pp. 369--380,
\newblock Springer, Dordrecht, 1993.

\bibitem{Thiel96}
{\sc W.~Thiel},
\newblock {\em Adv. Chem. Phys.} {\bf 93}, 703 (1996).

\bibitem{Thiel98}
{\sc W.~Thiel},
\newblock {\em Thermochemistry from Semiempirical Molecular Orbital Theory},
  chapter~8, pp. 142--161,
\newblock American Chemical Society: Washington, DC, 1998.

\bibitem{Clark00}
{\sc T.~Clark},
\newblock {\em J. Mol. Struct. THEOCHEM} {\bf 530}, 1 (2000).

\bibitem{Bredow05}
{\sc T.~Bredow} and {\sc K.~Jug},
\newblock {\em Theor. Chem. Acc.} {\bf 113}, 1 (2005).

\bibitem{Stewart07}
{\sc J.~J.~P. Stewart},
\newblock {\em Rev. Comput. Chem.} {\bf 1}, 45 (2007).

\bibitem{Lewars10}
{\sc E.~G. Lewars},
\newblock {\em Computational chemistry: Introduction to the theory and
  applications of molecular and quantum mechanics},
\newblock Springer Science \& Business Media, 2010.

\bibitem{Clark11}
{\sc T.~Clark} and {\sc J.~J.~P. Stewart},
\newblock {\em MNDO-Like Semiempirical Molecular Orbital Theory and Its
  Application to Large Systems}, chapter~8, pp. 259--286,
\newblock John Wiley \& Sons: New York, 2011.

\bibitem{Thiel14}
{\sc W.~Thiel},
\newblock {\em WIREs Comput Mol Sci} {\bf 4}, 145 (2014).

\bibitem{Bredow17}
{\sc T.~Bredow} and {\sc K.~Jug},
\newblock {\em Semiempirical Molecular Orbital Methods}, chapter~6, pp.
  159--202,
\newblock American Cancer Society, 2017.

\bibitem{Husch18}
{\sc T.~Husch}, {\sc A.~C. Vaucher}, and {\sc M.~Reiher},
\newblock {\em Int. J. Quantum Chem.} {\bf 118}, e25799 (2018).

\bibitem{Roothaan51}
{\sc C.~C.~J. Roothaan},
\newblock {\em Rev. Mod. Phys.} {\bf 23}, 69 (1951).

\bibitem{Hall51}
{\sc G.~G. Hall},
\newblock {\em Proc. R. Soc. Lond. A} {\bf 205}, 541 (1951).

\bibitem{Sun18}
{\sc Q.~Sun}, {\sc T.~C. Berkelbach}, {\sc N.~S. Blunt}, {\sc G.~H. Booth},
  {\sc S.~Guo}, {\sc Z.~Li}, {\sc J.~Liu}, {\sc J.~D. McClain}, {\sc E.~R.
  Sayfutyarova}, {\sc S.~Sharma}, {\sc S.~Wouters}, and {\sc G.~K.-L. Chan},
\newblock {\em WIREs Comput Mol Sci} {\bf 8}, e1340 (2018).

\bibitem{Weigend05}
{\sc F.~Weigend} and {\sc R.~Ahlrichs},
\newblock {\em Phys. Chem. Chem. Phys.} {\bf 7}, 3297 (2005).

\bibitem{Loewdin70}
{\sc P.-O. L\"owdin},
\newblock {\em Adv. Quantum Chem.} {\bf 5}, 185 (1970).

\bibitem{Givens58}
{\sc W.~Givens},
\newblock {\em SIAM J. Appl. Math.} {\bf 6}, 26 (1958).

\bibitem{Jacobi46}
{\sc C.~G.~J. Jacobi},
\newblock {\em Crelle's Journal} {\bf 30}, 51 (1846).

\bibitem{Robbe05}
{\sc M.~Robb{\'e}} and {\sc M.~Sadkane},
\newblock {\em BIT Numer. Math.} {\bf 45}, 181 (2005).

\bibitem{Cederbaum89}
{\sc L.~S. Cederbaum}, {\sc J.~Schirmer}, and {\sc H.-D. Meyer},
\newblock {\em J. Phys. A} {\bf 22}, 2427 (1989).

\bibitem{Sikkema09}
{\sc J.~Sikkema}, {\sc L.~Visscher}, {\sc T.~Saue}, and {\sc M.~Ilia\v{s}},
\newblock {\em J. Chem. Phys.} {\bf 131}, 124116 (2009).

\bibitem{Peng12}
{\sc D.~Peng} and {\sc M.~Reiher},
\newblock {\em J. Chem. Phys.} {\bf 136}, 244108 (2012).

\bibitem{Seino12}
{\sc J.~Seino} and {\sc H.~Nakai},
\newblock {\em J. Chem. Phys.} {\bf 136}, 244102 (2012).

\bibitem{Zilkowski09}
{\sc M.~Zi{\l}kowski}, {\sc B.~Jans{\'i}k}, {\sc P.~J{\o}rgensen}, and {\sc
  J.~Olsen},
\newblock {\em J. Chem. Phys.} {\bf 131}, 124112 (2009).

\bibitem{Li14}
{\sc Z.~Li}, {\sc H.~Li}, {\sc B.~Suo}, and {\sc W.~Liu},
\newblock {\em Acc. Chem. Res.} {\bf 47}, 2758 (2014).

\bibitem{Perdew86}
{\sc J.~P. Perdew},
\newblock {\em Phys. Rev. B} {\bf 33}, 8822 (1986).

\bibitem{Becke88}
{\sc A.~D. Becke},
\newblock {\em Phys. Rev. A} {\bf 38}, 3098 (1988).

\bibitem{Reiher09}
{\sc M.~Reiher} and {\sc A.~Wolf},
\newblock {\em Relativistic Quantum Chemistry: The Fundamental Theory of
  Molecular Science},
\newblock Wiley-VCH, Weinheim, 2nd edition, 2015.

\bibitem{Heully86}
{\sc J.-L. Heully}, {\sc I.~Lindgren}, {\sc E.~Lindroth}, {\sc S.~Lundquist},
  and {\sc A.-M. M{\aa}rtensen-Pendrill},
\newblock {\em J. Phys. B: At. Mol. Phys.} {\bf 19}, 2799 (1986).

\bibitem{Jensen05}
{\sc H.~J.~A. Jensen},
\newblock Talk on conference on relativistic effects in heavy elements ---
  REHE, April 2005, M\"ulheim, 2005.

\bibitem{Hess86}
{\sc B.~A. Hess},
\newblock {\em Phys. Rev. A} {\bf 33}, 3742 (1986).

\bibitem{wolf02b}
{\sc A.~Wolf}, {\sc M.~Reiher}, and {\sc B.~A. Hess},
\newblock {\em J. Chem. Phys.} {\bf 117}, 9215 (2002).

\bibitem{reih04}
{\sc M.~Reiher} and {\sc A.~Wolf},
\newblock {\em J. Chem. Phys.} {\bf 121}, 2037 (2004).

\bibitem{bary97}
{\sc M.~Barysz}, {\sc A.~J. Sadlej}, and {\sc J.~G. Snijders},
\newblock {\em Int. J. Quantum Chem.} {\bf 65}, 225 (1997).

\bibitem{bary01}
{\sc M.~Barysz} and {\sc A.~J. Sadlej},
\newblock {\em J. Mol. Struct. (THEOCHEM)} {\bf 573}, 181 (2001).

\bibitem{bary02}
{\sc M.~Barysz} and {\sc A.~J. Sadlej},
\newblock {\em J. Chem. Phys.} {\bf 116}, 2696 (2002).

\bibitem{fila03}
{\sc M.~Filatov} and {\sc D.~Cremer},
\newblock {\em J. Chem. Phys.} {\bf 119}, 11526 (2003).

\bibitem{fila03b}
{\sc M.~Filatov} and {\sc K.~G. Dyall},
\newblock {\em Theor. Chem. Acc.} {\bf 117}, 333 (2007).

\bibitem{reih04b}
{\sc M.~Reiher} and {\sc A.~Wolf},
\newblock {\em J. Chem. Phys.} {\bf 121}, 10945 (2004).

\bibitem{reih06}
{\sc M.~Reiher},
\newblock {\em Theor. Chem. Acc.} {\bf 116}, 241 (2006).

\bibitem{Kutzelnigg05}
{\sc W.~Kutzelnigg} and {\sc W.~Liu},
\newblock {\em J. Chem. Phys.} {\bf 123}, 241102 (2005).

\bibitem{Ilias07}
{\sc M.~Ilia\v{s}} and {\sc T.~Saue},
\newblock {\em J. Chem. Phys.} {\bf 126}, 064102 (2007).

\bibitem{Saue11}
{\sc T.~Saue},
\newblock {\em ChemPhysChem} {\bf 12}, 3077 (2011).

\bibitem{dyal97}
{\sc K.~G. Dyall},
\newblock {\em J. Chem. Phys.} {\bf 106}, 9618 (1997).

\bibitem{Liu06}
{\sc W.~Liu} and {\sc D.~Peng},
\newblock {\em J. Chem. Phys.} {\bf 125}, 044102 (2006).

\bibitem{Kutzelnigg06}
{\sc W.~Kutzelnigg} and {\sc W.~Liu},
\newblock {\em Mol. Phys.} {\bf 104}, 2225 (2006).

\bibitem{Liu07}
{\sc W.~Liu} and {\sc W.~Kutzelnigg},
\newblock {\em J. Chem. Phys.} {\bf 126}, 114107 (2007).

\bibitem{Peng12a}
{\sc D.~Peng} and {\sc M.~Reiher},
\newblock {\em Theor. Chem. Acc.} {\bf 131}, 1081 (2012).

\bibitem{Visscher97}
{\sc L.~Visscher},
\newblock {\em Theor. Chem. Acc.} {\bf 98}, 68 (1997).

\bibitem{mitt72}
{\sc M.~H. Mittleman},
\newblock {\em Phys. Rev. A} {\bf 5}, 2395 (1972).

\bibitem{mitt81}
{\sc M.~H. Mittleman},
\newblock {\em Phys. Rev. A} {\bf 24}, 1167 (1981).

\bibitem{such80}
{\sc J.~Sucher},
\newblock {\em Phys. Rev. A} {\bf 22}, 348 (1980).

\bibitem{such87}
{\sc J.~Sucher},
\newblock {\em Phys. Scr.} {\bf 36}, 271 (1987).

\bibitem{hess96}
{\sc B.~A. He{\ss}}, {\sc C.~M. Marian}, {\sc U.~Wahlgren}, and {\sc
  O.~Gropen},
\newblock {\em Chem. Phys. Lett.} {\bf 251}, 365 (1996).

\bibitem{schi98b}
{\sc B.~Schimmelpfennig}, {\sc L.~Maron}, {\sc U.~Wahlgren}, {\sc
  C.~Teichteil}, {\sc H.~Fagerli}, and {\sc O.~Gropen},
\newblock {\em Chem. Phys. Lett.} {\bf 286}, 261 (1998).

\bibitem{Autschbach12}
{\sc J.~Autschbach}, {\sc D.~Peng}, and {\sc M.~Reiher},
\newblock {\em J. Chem. Theory Comput.} {\bf 8}, 4239 (2012).

\bibitem{Peralta04}
{\sc J.~E. Peralta} and {\sc G.~E. Scuseria},
\newblock {\em J. Chem. Phys.} {\bf 120}, 5875 (2004).

\bibitem{Peralta05}
{\sc J.~E. Peralta}, {\sc J.~Uddin}, and {\sc G.~E. Scuseria},
\newblock {\em J. Chem. Phys.} {\bf 122}, 084108 (2005).

\bibitem{Thar09}
{\sc J.~Thar} and {\sc B.~Kirchner},
\newblock {\em J. Chem. Phys.} {\bf 130}, 124103 (2009).

\end{thebibliography}

\end{document}